\documentclass[a4paper]{article}

\usepackage{a4wide}
\usepackage[UKenglish]{babel}
\usepackage{graphicx,subfigure}
	\graphicspath{{./}{figures/}}
\usepackage[colorlinks=true,linkcolor=blue,citecolor=red]{hyperref}
\usepackage{xcolor}
\usepackage{amssymb,amsthm,empheq,bbold}
\usepackage[inline]{enumitem}
\usepackage[ruled,vlined,linesnumbered]{algorithm2e}
\usepackage{caption} 
\usepackage[normalem]{ulem}

\theoremstyle{remark}\newtheorem{remark}{Remark}[section]

\newcommand{\abs}[1]{\left\lvert#1\right\rvert}
\DeclarePairedDelimiter\ave{\langle}{\rangle}
\newcommand{\cI}{\mathcal{I}}
\renewcommand{\P}{\operatorname{Prob}}
\newcommand{\R}{\mathbb{R}}

\newcommand{\Var}{\operatorname{Var}}

\allowdisplaybreaks

\begin{document}
\title{Boltzmann-type equations for multi-agent systems with label switching}
\author{Nadia Loy\thanks{\texttt{nadia.loy@polito.it}} \and Andrea Tosin\thanks{\texttt{andrea.tosin@polito.it}}}
\date{\small Department of Mathematical Sciences ``G. L. Lagrange'' \\ Politecnico di Torino, Italy}

\maketitle

\begin{abstract}
In this paper, we propose a Boltzmann-type kinetic description of mass-varying interacting multi-agent systems. Our agents are characterised by a microscopic state, which changes due to their mutual interactions, and by a label, which identifies a group to which they belong. Besides interacting within and across the groups, the agents may change label according to a state-dependent Markov-type jump process. We derive general kinetic equations for the joint interaction/label switch processes in each group. For prototypical birth/death dynamics, we characterise the transient and equilibrium kinetic distributions of the groups via a Fokker-Planck asymptotic analysis. Then we introduce and analyse a simple model for the contagion of infectious diseases, which takes advantage of the joint interaction/label switch processes to describe quarantine measures.

\medskip


\medskip

\end{abstract}

\section{Introduction}
Boltzmann-type kinetic equations are a valuable tool to model multi-agent systems, their success being confirmed by a great variety of modern applications which take advantage of the formalism of the collisional kinetic theory. The literature in the field is constantly growing as witnessed by very recent contributions to econophysics~\cite{dimarco2020PRE}, human ecology~\cite{toscani2019PRE}, vehicular traffic with autonomous vehicles~\cite{piccoli2020ZAMP,tosin2019MMS,tosin2021MCRF}, opinion formation~\cite{pareschi2019JNS}, biology~\cite{loy2019JMB,loy2020KRM,preziosi2021JTB}.

Kinetic models of multi-agent systems are based on a revisitation of the methods of the classical kinetic theory, however with remarkable differences due to the different nature of the physical systems at hand. Classical kinetic theory deals mostly with the dynamics of gas molecules and their elastic collisions, which conserve microscopically both the momentum and the kinetic energy of the pairs of colliding molecules. Instead, interactions in multi-agent systems often do not conserve either the first or the second statistical moment of the distribution function, which does not only have consequences on the physical interpretation of the dynamics but also on the techniques required to investigate it mathematically. On the other hand, the virtually ubiquitous characteristic of kinetic models of multi-agent systems is the fact that the total number of agents does not change in time. This is equivalent to the conservation of mass in the collisions among gas molecules and allows one to regard the kinetic distribution function as a probability density function. Then, the kinetic equations may be derived from stochastic microscopic interaction dynamics by appealing to probabilistic arguments.

In this paper, we are instead interested in multi-agent dynamics which do not conserve necessarily the number of agents. Indeed, many applications in population dynamics involve processes assimilable to ``birth'' and ``death'', i.e. to the appearance and disappearance of interacting agents on the basis of the microscopic state that they are currently expressing. A similar concept in classical applications of kinetic theory to gas and plasma dynamics is the absorption of particles, see e.g.,~\cite{cercignani1988BOOK}. In particular, we focus on a \textit{label switch} process, that we may summarise as follows: while interacting, agents may change a label which denotes their membership of a particular group/category within the whole population. As a consequence, the number of agents in each group varies in time. This process clearly affects the interactions in each group, both because the number of interacting agents changes and because the distribution of the microscopic states in each group is altered by the introduction or the removal of agents.

Our label switch process is conceptually analogous to chemical reactions in which the total density of the molecules is conserved in time whereas that of each species is not. This type of problems has been deeply studied in the kinetic literature, see e.g.,~\cite{groppi2004JSP,groppi1999JMC,moreau1975PHYSA,rossani1999PHYSA}. In our case, however, the strongly different physical nature of the particles under consideration and of their microscopic dynamics requires to elaborate a different form of the kinetic equations describing the label switching. Recently, other contributions dealing with label switching in particle systems have been proposed, see e.g.,~\cite{albi2019M3AS,morandotti2020SIMA}. While sharing some conceptual analogies with our microscopic agent dynamics, these works focus on mean-field descriptions, which in principle may be regarded as particular cases of Boltzmann-type kinetic descriptions. Indeed, in some regimes Boltzmann-type ``collisional'' models might be approximated by mean-field models but in general they contain a much richer variety of trends depending on the ranges of the parameters. It is therefore interesting to investigate systematically a Boltzmann-type approach in presence of label switching. This is the aim and the main novelty of this work compared to other studies about similar topics.

In more detail, the paper is organised as follows: in Section~\ref{sect:preliminaries} we review separately some basic facts about the Boltzmann-type description of binary interactions and Markov-type jump processes, which in this context provide a proper mathematical framework to formalise the label switch process. In Section~\ref{sect:int+labswitch} we derive the Boltzmann-type equations with binary interactions and label switching starting from a stochastic microscopic description of the interaction and relabelling processes. Within this formalism, in Section~\ref{sect:death.birth} we investigate the transient and asymptotic trends of birth and death processes of interacting agents, which can be regarded as the prototypes of a wide range of non-conservative particle dynamics. In Section~\ref{sect:contagion} we propose an application of the mathematical structures previously developed to a problem of contagion of infectious diseases with quarantine. Finally, in Section~\ref{sect:numerical} we show some numerical simulations of the kinetic model of Section~\ref{sect:contagion} obtained by means of a Monte Carlo particle algorithm that we derive straightforwardly from the stochastic microscopic description introduced in Section~\ref{sect:int+labswitch}. The algorithm is reported in Appendix~\ref{app:nanbu}.

\section{Preliminaries on labelled interacting agents}
\label{sect:preliminaries}
Let us consider a large system of agents described by a microscopic state $v\in\R_+$ representing a non-negative physical quantity. Extensions to negative and possibly also bounded microscopic states are mostly a matter of technicalities using the very same ideas presented in this paper. The agents may belong to different groups or categories identified by a discrete label $x\in\cI=\{1,\,\dots,\,n\}$, that the agents may change as a result of a Markov-type jump process. Such a stochastic \textit{label switch} process is defined by a transition probability
\begin{equation}
	T=T(v,t;\,x\vert y)\in [0,\,1] \qquad \forall\,v\in\R_+,\ x,\,y\in\cI,\ t>0,
	\label{trans_prob}
\end{equation}  
namely the probability that an agent with state $v$ at time $t$ switches from label $y$ to label $x$. In order for $T(v,t;\,\cdot\vert y)$ to be a conditional probability density, it has to satisfy the following further property:
$$ \int_\cI T(v,t;\,x\vert y)\,dx=1 \qquad \forall\,v\in\R_+,\ y\in\cI,\ t>0. $$
A label switch corresponds therefore to a migration of an agent to a different group, however in such a way that the total mass of the agents in the system is conserved. We say that this process is formally a Markov-type one because the probability to switch from the current label $y$ to a new label $x$ does not depend on how the agent reached previously the label $y$.

\begin{remark} \label{rem:discrete}
Since the variable $x$ is discrete, the mapping $x\mapsto T(v,t;\,x\vert y)$ is a discrete probability measure. Consequently, we actually have
$$ \int_\cI T(v,t;\,x\vert y)\,dx=\sum_{i=1}^{n}T(v,t;\,i\vert y). $$
\end{remark}

Agents within the same group, i.e. with the same label, are assumed to be indistinguishable. Their microscopic state $v$ evolves in consequence of \textit{binary interactions} with either other agents of the same group or agents belonging to a different group. We will take into account the possibility that the interactions among agents with the same label differ from those among agents with different labels. In general, if $v,\,v_\ast\in\R_+$ denote the pre-interaction states of any two interacting agents, their post-interaction states $v',\,v_\ast'$ will be given by general \textit{linear} microscopic rules of the form
\begin{equation}
	v'=p_1v+q_1v_\ast, \qquad v_\ast'=p_2v_\ast+q_2v,
	\label{eq:binary_gen}
\end{equation}
where $p_i,\,q_i\in\R_+$ for $i=1,\,2$ are either deterministic or stochastic coefficients. As a particular relevant sub-case, we will consider \textit{symmetric} interactions, namely those with $p_1=p_2$ and $q_1=q_2$. In this case, it will be generally sufficient to refer only to the representative rule
\begin{equation}
	v'=pv+qv_\ast
	\label{eq:binary_gen_sym}
\end{equation}
with $p,\,q\in\R_+$ independent random variables.

\subsection{Boltzmann-type description of the interaction dynamics}
\label{sect:int.dyn}
It is known that an aggregate description of the (sole) interaction dynamics~\eqref{eq:binary_gen_sym} inspired by the principles of statistical mechanics can be obtained by introducing a distribution function $f=f(v,t)\geq 0$ such that $f(v,t)dv$ gives the proportion of agents having at time $t$ a microscopic state comprised between $v$ and $v+dv$. Such a distribution function satisfies a Boltzmann-type kinetic equation, which in weak form reads
\begin{equation}
	\frac{d}{dt}\int_{\R_+}\varphi(v)f(v,t)\,dv=\mu\int_{\R_+}\int_{\R_+}\ave{\varphi(v')-\varphi(v)}f(v,t)f(v_\ast,t)\,dv\,dv_\ast,
	\label{eq:boltz}
\end{equation}
where $\varphi:\R_+\to\R$ is an observable quantity (test function) and $\mu>0$ is the interaction frequency, here assumed to be constant. Moreover, $\ave{\cdot}$ denotes expectation with respect to the laws of the stochastic coefficients $p,\,q$. This equation expresses the fact that the time variation of the expectation of $\varphi$ (left-hand side) is due to the mean variation of $\varphi$ in a binary interaction (right-hand side). For a detailed derivation of~\eqref{eq:boltz} we refer the interested reader to~\cite{pareschi2013BOOK}.

Choosing $\varphi(v)=1$ we obtain
$$ \frac{d}{dt}\int_{\R_+} f(v,t)\,dv=0, $$
which means that the total mass of the agents is conserved in time by the interactions~\eqref{eq:binary_gen_sym}.

Choosing instead $\varphi(v)=v^n$, $n=1,\,2,\,\dots$, we obtain the evolution of the statistical moments of $f$. For instance, with $\varphi(v)=v$ we find that the mean $m(t):=\int_{\R_+}vf(v,t)\,dv$ evolves according to
$$ \frac{dm}{dt}=\ave{p+q-1}m. $$
If $\ave{p+q}=1$ then $m$ is conserved in time while if $\ave{p+q}>1$ or $\ave{p+q}<1$ then $m$ either blows to infinity or decreases to zero exponentially fast in time.

The trend of the total energy $E(t):=\int_{\R_+}v^2f(v,t)\,dv$ is obtained from $\varphi(v)=v^2$ and turns out to be ruled by the equation
$$ \frac{dE}{dt}=\ave{p^2+q^2-1}E+2\ave{pq}m^2. $$

From the trend of $m$ and $E$ we can also infer that of the internal energy $e(t):=E(t)-m^2(t)$, namely the variance of the distribution $f$:
$$ \frac{de}{dt}=\ave{p^2+q^2-1}e+\ave{(p+q-1)^2}m^2. $$

Notice that if $m$ is conserved in time and $m\neq 0$ then both $E$ and $e$ tend asymptotically to a finite non-vanishing value if $\ave{p^2+q^2}<1$.

\subsection{Kinetic description of the label switching}
\label{sect:label.switch}
If we consider only the label switch process then the evolution of the distribution function $f=f(x,t)\geq 0$ of the agents with label $x$ at time $t$ can be modelled by a kinetic equation describing a Markov-type jump process, see~\cite{loy2020CMS}:
\begin{equation}
	\partial_tf(x,t)=\lambda\left(\int_\cI T(t;\,x\vert y)f(y,t)\,dy-f(x,t)\right),
	\label{eq:mark_proc_1}
\end{equation}
where $\lambda>0$ is the (constant) switch rate. In weak form~\eqref{eq:mark_proc_1} reads
\begin{equation}
	\frac{d}{dt}\int_\cI\psi(x)f(x,t)\,dx=\lambda\int_\cI\int_\cI(\psi(x)-\psi(y))T(t;\,x\vert y)f(y,t)\,dx\,dy,
	\label{eq:jump_proc_weak}
\end{equation}
where $\psi:\cI\to\R$ is another observable quantity (test function).

Since $x\in\cI$ is discrete, we may conveniently represent the distribution function $f$ as
\begin{equation}
	f(x,t)=\sum_{i=1}^{n}f_i(t)\delta(x-i),
	\label{eq:f.delta.label_switch}
\end{equation}
where $\delta(x-i)$ is the Dirac distribution centred in $x=i$ and $f_i=f_i(t)\geq 0$ is the probability that an agent is labelled by $x=i$ at time $t$. In this way, we reconcile the weak form~\eqref{eq:jump_proc_weak} with the convention introduced in Remark~\ref{rem:discrete}, for~\eqref{eq:jump_proc_weak} actually becomes
$$ \sum_{i=1}^{n}\psi(i)f_i'(t)=\lambda\sum_{i=1}^{n}\sum_{j=1}^{n}(\psi(i)-\psi(j))T(t;\,i\vert j)f_j(t). $$
Choosing $\psi$ such that $\psi(i)=1$ for a certain $i\in\cI$ and $\psi(x)=0$ for all $x\in\cI\setminus\{i\}$ we get in particular
\begin{equation}
	f_i'=\lambda\left(\sum_{j=1}^{n}T(t;\,i\vert j)f_j-f_i\right), \qquad i=1,\,\dots,\,n.
	\label{eq:fi.jump}
\end{equation}

\section{Kinetic description of interactions with label switching}
\label{sect:int+labswitch}
We now want to derive a kinetic equation for the joint distribution function $f=f(x,v,t)\geq 0$, such that $f(x,v,t)dv$ gives the proportion of agents labelled by $x\in\cI$ and having microscopic state comprised between $v$ and $v+dv$ at time $t$. The discreteness of $x$ allows us to represent $f$ as
$$ f(x,v,t)=\sum_{i=1}^{n}f_i(v,t)\delta(x-i), $$
where $f_i=f_i(v,t)\geq 0$ is the distribution function of the microscopic state $v$ of the agents with label $i$ and, in particular, $f_i(v,t)dv$ is the proportion of agents with label $i$ whose microscopic state is comprised between $v$ and $v+dv$ at time $t$.

Since both the interactions and the label switching conserve the total mass of the system, we may assume that $f(x,v,t)$ is a probability distribution, namely:
\begin{equation}
	\int_{\R_+}\int_\cI f(x,v,t)\,dx\,dv=\sum_{i=1}^{n}\int_{\R_+} f_i(v,t)\,dv=1 \qquad \forall\,t>0.
	\label{eq:f.prob}
\end{equation}
Notice, however, that the $f_i$'s are in general not probability density functions because their $v$-integral varies in time due to the label switching. We denote by
\begin{equation}
	\rho_i(t):=\int_{\R_+} f_i(v,t)\,dv
	\label{eq:rhoi}
\end{equation}
the mass of the group of agents with label $i$, thus $0\leq\rho_i(t)\leq 1$ and
$$ \sum_{i=1}^{n}\rho_i(t)=1 \qquad \forall\,t>0. $$

The following derivation is an extension of the one introduced in~\cite{pareschi2013BOOK}. We perform it in some detail because it provides also a formal justification of a Monte Carlo method for the numerical solution of the resulting kinetic equation. We will use this method later in Section~\ref{sect:numerical} for our numerical tests.

Let $(X_t,\,V_t)\in\cI\times\R_+$ be a pair of random variables denoting the label and the microscopic state of a representative agent of the system at time $t$. The joint probability distribution of such a pair is $f(x,v,t)$. During a sufficiently small time $\Delta{t}>0$ the agent may or may not change the pair $(X_t,\,V_t)$ depending on whether a label switch and/or a binary interaction with another agent takes place. We express this discrete-in-time random process as
\begin{align}
	\begin{aligned}[c]
		X_{t+\Delta{t}} &= (1-\Theta)X_t+\Theta J_t, \\
		V_{t+\Delta{t}} &= (1-\Xi)V_t+\Xi V'_t,
	\end{aligned}
	\label{eq:micro.rules.gen}
\end{align} 
where $J_t$, $V'_t$ are random variables describing the new label after a label switch and the new microscopic state after a binary interaction, respectively, while $\Theta,\,\Xi\in\{0,\,1\}$ are Bernoulli random variables, which we assume independent of all the other variables appearing in~\eqref{eq:micro.rules.gen}, discriminating whether a label switch and a binary interaction take place ($\Theta,\,\Xi=1$) or not ($\Theta,\,\Xi=0$) during the time $\Delta{t}$. In particular, we set
\begin{equation}
	\P(\Theta=1)=\lambda\Delta{t}, \qquad \P(\Xi=1)=\mu\Delta{t}
	\label{eq:bernoulli}
\end{equation}
where $\lambda$, $\mu$ are the frequencies introduced in Sections~\ref{sect:int.dyn},~\ref{sect:label.switch} and $\Delta{t}\leq\min\{\frac{1}{\lambda},\,\frac{1}{\mu}\}$ for consistency. The underlying assumption is that the longer the time interval $\Delta{t}$ the higher the probability that a label switch and/or a binary interaction takes place. Notice that $\frac{1}{\lambda}$, $\frac{1}{\mu}$ can be understood as the mean waiting times between two successive label switches/binary interactions, respectively.

The random variable $J_t\in\cI$ models the Markov-type jump process leading to a label switch. If $P(j,v,t)$ denotes the joint probability distribution of the pair $(J_t,\,V_t)$ then
$$ P(j,v,t)=\int_\cI T(v,t;\,j\vert y)f(y,v,t)\,dy, $$
where $T(v,t;\,j\vert y)$ is the transition probability~\eqref{trans_prob}.

The random variable $V'_t\in\R_+$ gives instead the new microscopic state after a binary interaction with another agent described by the pair $(X^\ast_t,\,V^\ast_t)\in\cI\times\R_+$. In order to account for possibly different interaction rules depending on the labels of the interacting agents, we define
\begin{equation}
	V'_t:=\delta_{X_t,X^\ast_t}\bar{V}'_t+(1-\delta_{X_t,X^\ast_t})\tilde{V}'_t
	\label{eq:micro.rule_exch}
\end{equation}
where
$$	\delta_{X,X^\ast}=
	\begin{cases}
		1 & \text{if } X=X^\ast \\
		0 & \text{if } X\neq X^\ast
	\end{cases} $$
is the Kronecker delta. In particular, $\bar{V}'_t,\,\tilde{V}'_t\in\R_+$ represent the outcomes of binary interactions between agents with the same and different labels, respectively. They will be both of the form~\eqref{eq:binary_gen_sym}, namely
$$ \bar{V}'_t=\bar{p}V_t+\bar{q}V^\ast_t, \qquad \tilde{V}'_t=\tilde{p}V_t+\tilde{q}V^\ast_t $$
with $\bar{p},\,\bar{q},\,\tilde{p},\,\tilde{q}\in\R$ either deterministic or independent random coefficients.

Let now $\phi=\phi(x,v)$ be an observable quantity defined on $\cI\times\R_+$. From~\eqref{eq:micro.rules.gen},~\eqref{eq:bernoulli}, together with the assumed independence of $\Theta,\,\Xi$, we see that the mean variation rate of $\phi$ in the time interval $\Delta{t}$ satisfies
\begin{align*}
	&\frac{\ave{\phi(X_{t+\Delta{t}},V_{t+\Delta{t}})}-\ave{\phi(X_t,V_t)}}{\Delta{t}}= \\
	&\qquad\qquad\phantom{+} \frac{(1-\lambda\Delta{t})(1-\mu\Delta{t})\ave{\phi(X_t,V_t)}+\mu\Delta{t}(1-\lambda\Delta{t})\ave{\phi(X_t,V'_t)}}{\Delta{t}} \\
	&\qquad\qquad +\dfrac{\lambda\Delta{t}(1-\mu\Delta{t})\ave{\phi(J_t,V_t)}+\lambda\mu{\Delta{t}}^2\ave{\phi(J_t,V'_t)}-\ave{\phi(X_t,V_t)}}{\Delta{t}},
\end{align*}
whence we deduce the instantaneous time variation of the average of $\phi$ in the limit $\Delta{t}\to 0^+$ as
$$ \frac{d}{dt}\ave{\phi(X_t,V_t)}=\lambda\ave{\phi(J_t,V_t)}+\mu\ave{\phi(X_t,V'_t)}-(\lambda+\mu)\ave{\phi(X_t,V_t)}. $$
Notice that the simultaneous change of label and microscopic state, i.e. the term $\ave{\phi(J_t,V'_t)}$, turns out to be a higher order effect in time disregarded in this limit equation. Owing to~\eqref{eq:micro.rule_exch}, we further obtain
\begin{align}
	\begin{aligned}[b]
		\frac{d}{dt}\ave{\phi(X_t,V_t)} &= \lambda\ave{\phi(J_t,V_t)} \\
		&\phantom{=} +\mu\ave{\delta_{X_t,X^\ast_t}\phi(X_t,\bar{V}'_t)}+\mu\ave{(1-\delta_{X_t,X^\ast_t})\phi(X_t,\tilde{V}'_t)} \\
		&\phantom{=} -(\lambda+\mu)\ave{\phi(X_t,V_t)}.
	\end{aligned}
	\label{eq:mean_obs}
\end{align}

We consider now that
$$ \ave{\phi(J_t,V_t)}=\int_{\R_+}\int_{\cI}\phi(i,v)P(i,v,t)\,di\,dv=\sum_{i=1}^{n}\sum_{j=1}^{n}\int_{\R_+}\phi(i,v)T(v,t;\,i|j)f_j(v,t)\,dv $$
and that
\begin{align*}
	\ave{\delta_{X_t,X^\ast_t}\phi(X_t,\bar{V}'_t)} &= \sum_{i=1}^{n}\int_{\R_+}\int_{\R_+}\ave{\phi(i,\bar{v}')}f_i(v,t)f_i(v_\ast,t)\,dv\,dv_\ast \\
	\ave{(1-\delta_{X_t,X^\ast_t})\phi(X_t,\tilde{V}'_t)} &= \sum_{i=1}^{n}\sum_{\substack{j=1 \\ j\neq i}}^{n}\int_{\R_+}\int_{\R_+}\ave{\phi(i,\tilde{v}')}f_i(v,t)f_j(v_\ast,t)\,dv\,dv_\ast,
\end{align*}
where $\ave{\cdot}$ denotes the average with respect to the possibly random coefficients $\bar{p}$, $\bar{q}$, $\tilde{p}$, $\tilde{q}$ contained in $\bar{v}'$ and $\tilde{v}'$. As typically done in kinetic theory~\cite{cercignani1988BOOK}, in writing these interaction terms we assume the \textit{propagation of chaos}, which allows us to perform the factorisation $f(x,v,y,v_\ast,t)=f(x,v,t)f(y,v_\ast,t)$ of the two-particle distribution function. Hence from~\eqref{eq:mean_obs} we deduce the following equation:
\begin{align*}
	\frac{d}{dt}\sum_{i=1}^{n}\int_{\R_+}\phi(i,v)f_i(v,t)\,dv &= \lambda\sum_{i=1}^{n}\sum_{j=1}^{n}\int_{\R_+}\phi(i,v)T(v,t;\,i|j)f_j(v,t)\,dv \\
	&\phantom{=} +\mu\sum_{i=1}^{n}\int_{\R_+}\int_{\R_+}\ave{\phi(i,\bar{v}')}f_i(v,t)f_i(v_\ast,t)\,dv\,dv_\ast \\
	&\phantom{=} +\mu\sum_{i=1}^{n}\sum_{\substack{j=1 \\ j\neq i}}^{n}\int_{\R_+}\int_{\R_+}\ave{\phi(i,\tilde{v}')}f_i(v,t)f_j(v_\ast,t)\,dv\,dv_\ast \\
	&\phantom{=} -(\lambda+\mu)\sum_{i=1}^{n}\int_{\R_+}\phi(i,v)f_i(v,t)\,dv,
\end{align*}
which has to hold for every $\phi:\cI\times\R_+\to\R$. Choosing $\phi(x,v)=\psi(x)\varphi(v)$ with $\psi$ such that $\psi(i)=1$ for a certain $i\in\cI$ and $\psi(x)=0$ for all $x\in\cI\setminus\{i\}$ and exploiting~\eqref{eq:f.prob} to merge the loss term (last term on the right-hand side) with the other terms on the right-hand side, we finally obtain the following system of equations for the $f_i$'s:
\begin{align}
	\begin{aligned}[b]
		\frac{d}{dt}\int_{\R_+}\varphi(v)f_i(v,t)\,dv &= \lambda\int_{\R_+}\varphi(v)\left(\sum_{j=1}^{n}T(v,t;\,i\vert j)f_j(v,t)-f_i(v,t)\right)dv \\
		&\phantom{=} +\mu\int_{\R_+}\int_{\R_+}\ave{\varphi(\bar{v}')-\varphi(v)}f_i(v,t)f_i(v_\ast,t)\,dv\,dv_\ast \\
		&\phantom{=} +\mu\sum_{\substack{j=1 \\ j\neq i}}^{n}\int_{\R_+}\int_{\R_+}\ave{\varphi(\tilde{v}')-\varphi(v)}f_i(v,t)f_j(v_\ast,t)\,dv\,dv_\ast, \quad i=1,\,\dots,\,n.
	\end{aligned}
	\label{eq:boltz.fi}
\end{align}

Letting $\varphi(v)=1$ we discover that the mass $\rho_i$ of the agents with label $i$, cf.~\eqref{eq:rhoi}, evolves according to
$$ \frac{d\rho_i}{dt}+\lambda\rho_i=\lambda\sum_{j=1}^{n}\int_{\R_+}T(v,t;\,i\vert j)f_j(v,t)\,dv, $$
which depends explicitly on the label switch process.

As a particular case, we may reproduce in~\eqref{eq:boltz.fi} the situation in which only agents with the same label interact by letting $\tilde{v}'=v$. This corresponds to saying that interactions among agents with different labels do not produce a change of microscopic state, hence they are actually ``non-interactions''. Consequently,~\eqref{eq:boltz.fi} simplifies as
\begin{align}
	\begin{aligned}[b]
		\frac{d}{dt}\int_{\R_+}\varphi(v)f_i(v,t)\,dv &= \lambda\int_{\R_+}\varphi(v)\left(\sum_{j=1}^{n}T(v,t;\,i\vert j)f_j(v,t)-f_i(v,t)\right)dv \\
		&\phantom{=} +\mu\int_{\R_+}\int_{\R_+}\ave{\varphi(v')-\varphi(v)}f_i(v,t)f_i(v_\ast,t)\,dv\,dv_\ast, \quad i=1,\,\dots,\,n
	\end{aligned}
	\label{eq:boltz.fi-same_label}
\end{align}
with $v'$ given e.g., by~\eqref{eq:binary_gen_sym}.

\section{Death and birth processes}
\label{sect:death.birth}
We now use the kinetic equations derived in the previous section to study \textit{death and birth processes}. We regard them as prototypes and building blocks of general label switch dynamics in a wide range of applications, one of which will be illustrated in the next Section~\ref{sect:contagion}. Kinetic and mean-field descriptions of birth-death processes have been considered also in the recent literature, see e.g.,~\cite{albi2019M3AS,greenman2016PRE}. We mention moreover the kinetic approach to \textit{growth processes} described in~\cite{bassetti2015M3AS} and in~\cite[Chapter 2]{pareschi2013BOOK}, where, unlike our case, the size itself of the ensemble of agents is taken as the microscopic state of the system and the evolution of the corresponding probability density function is modelled. In the present context, the interest is due to the fact that the structure of our kinetic equations allows for a quite detailed characterisation of the transient and equilibrium distribution functions of the various groups of agents, possibly via suitable asymptotic analyses.

To be definite, we consider $n=2$ labels: $i=1$ denotes interacting or ``living'' agents whilst $i=2$ denotes inert or ``dead'' agents. The total mass of the agents is conserved but the mass of the agents with either label may change in time in consequence of label switches, i.e. ``deaths'' (transitions from $i=1$ to $i=2$) or ``births'' (transitions from $i=2$ to $i=1$).

Since we consider the agents labelled with $i=2$ as inert, we implicitly mean that they do not interact either with one another or with the agents labelled with $i=1$. Therefore, the reference equation for this application is~\eqref{eq:boltz.fi-same_label} with $v'=v$ for $i=2$.

\subsection{Death}
We begin by considering the death process only, in which only transitions from $i=1$ to $i=2$ are possible. Therefore, the transition probabilities describing the Markov-type jump process may be chosen as
\begin{equation}
	\begin{array}{ll}
		T(v,t;\,1\vert 2)=0, & T(v,t;\,2\vert 2)=1 \\
		T(v,t;\,2\vert 1)=\beta(v,t), & T(v,t;\,1\vert 1)=1-\beta(v,t)
	\end{array}
	\label{eq:trans_prob.death}
\end{equation}
with $0\leq\beta(v,t)\leq 1$ for all $v\in\R_+$ and $t>0$. From~\eqref{eq:boltz.fi-same_label}, the evolution equations for the distribution functions $f_1$, $f_2$ take then the form
\begin{align}
	\begin{aligned}[b]
		\frac{d}{dt}\int_{\R_+}\varphi(v)f_1(v,t)\,dv &= -\lambda\int_{\R_+}\varphi(v)\beta(v,t)f_1(v,t)\,dv \\
		&\phantom{=} +\mu\int_{\R_+}\int_{\R_+}\ave{\varphi(v')-\varphi(v)}f_1(v,t)f_1(v_\ast,t)\,dv\,dv_\ast
	\end{aligned}
	\label{eq:death_f1_weak}
\end{align}
and
\begin{equation}
	\frac{d}{dt}\int_{\R_+}\varphi(v)f_2(v,t)\,dv=\lambda\int_{\R_+}\varphi(v)\beta(v,t)f_1(v,t)\,dv.
	\label{eq:death_f2_weak}
\end{equation}

\subsubsection{Mass balance}
\label{sect:death.mass}
Letting $\varphi(v)=1$ in~\eqref{eq:death_f1_weak},~\eqref{eq:death_f2_weak} yields the time evolution of the masses $\rho_1$, $\rho_2$ of the two groups of agents:
\begin{align}
	& \frac{d\rho_1}{dt}=-\lambda\int_{\R_+}\beta(v,t)f_1(v,t)\,dv  \label{eq:death.mass_balance} \\
	& \frac{d\rho_2}{dt}=\lambda\int_{\R_+}\beta(v,t)f_1(v,t)\,dv. \nonumber
\end{align}
Actually, since $\rho_1(t)+\rho_2(t)$ is constant, if we assume a unitary total mass we may replace the second equation simply by $\rho_2(t)=1-\rho_1(t)$.

If the transition probability $\beta$ does not depend on $v$, i.e. $\beta=\beta(t)$, then we get in particular
$$ \rho_1(t)=\rho_{1,0}\exp{\left(-\lambda\int_0^t\beta(s)\,ds\right)}, $$
$\rho_{1,0}\in [0,\,1]$ being the prescribed mass $\rho_1$ at the initial time $t=0$. From here we see that $\rho_1$ tends to vanish asymptotically in time if e.g., $\beta$ does not depend on $t$ or if it approaches a constant non-zero value for large times. Conversely, if $\beta$ vanishes definitively from a certain time $t=t_0$ on then a residual mass of agents with label $i=1$ remains for large times.

If $\beta$ features a full dependence on $v$ and $t$, we cannot deduce from~\eqref{eq:death.mass_balance} an explicit expression for $\rho_1(t)$. Nevertheless, we observe that if there exists $\beta_0>0$ such that  $\beta(v,t)\geq\beta_0$ for all $v\in\R_+$ and all $t>0$ then 
$$ \rho_1(t)\leq\rho_{1,0}e^{-\lambda\beta_0 t}, $$
which implies that $\rho_1$ still vanishes for $t\to +\infty$. Consequently, we deduce $f_1(\cdot,t)\to 0$ in $L^1(\R_+)$ for $t\to +\infty$.

\subsubsection{Quasi-invariant limit and asymptotic distributions}
\label{sect:death.quasi-invariant_limit}
One of the most interesting issues in the study of kinetic models is the characterisation of the stationary distributions arising asymptotically for $t\to +\infty$, which depict the emergent behaviour of the system. For conservative kinetic equations this is typically carried out by means of asymptotic procedures, which, in suitable regimes of the parameters of the microscopic interactions, transform a Boltzmann-type integro-differential equation into a partial differential equation usually more amenable to analytical investigations. An effective asymptotic procedure is the so-called \textit{quasi-invariant limit}, which leads to \textit{Fokker-Planck-type} equations.

The idea behind the quasi-invariant limit is that one studies a regime in which the post-interaction state $v'$ is close enough to the pre-interaction state $v$, so that interactions produce a small transfer of microscopic state between the interacting agents. This concept was first introduced in the kinetic literature on multi-agent systems in~\cite{cordier2005JSP,toscani2006CMS} for binary collisions and in~\cite{furioli2017M3AS} for the interactions with a fixed background and has its roots in the concept of \textit{grazing collisions} studied in the classical kinetic theory~\cite{villani1998ARMA}.

In the present context, we extend this procedure to quasi-invariant microscopic dynamics encompassing both quasi-invariant interactions and quasi-invariant transition probabilities. We anticipate that, as a result, we obtain Fokker-Planck equations \textit{with reaction terms} linked to the label switch process. Interestingly, these equations are explicitly solvable at least in some representative cases, whereby analytical approximations of the Maxwellians can be derived.

In~\eqref{eq:binary_gen_sym}, after introducing a small parameter $0<\epsilon\ll 1$, we scale the coefficients as $p\to p^\epsilon$, $q\to q^\epsilon$, where $p^\epsilon$, $q^\epsilon$ are random variables such that
\begin{equation}
	\begin{array}{ll}
		\ave{p^\epsilon}=1-\epsilon, & \Var(p^\epsilon)=\kappa\epsilon \\[2mm]
		\ave{q^\epsilon}=\epsilon, & \Var(q^\epsilon)=\kappa\epsilon^{1+\theta}
	\end{array}
	\label{eq:quasi-invariant_scaling}
\end{equation}
and $\kappa,\,\theta>0$ are constant parameters. These choices are motivated by the following considerations: for $\epsilon\to 0^+$, on one hand $p^\epsilon$, $q^\epsilon$ converge in law to the constants $p=1$, $q=0$, respectively, thus in the regime of small $\epsilon$ the interaction~\eqref{eq:binary_gen_sym} is quasi-invariant. On the other hand, for finite $\epsilon>0$ it results $\ave{p^\epsilon+q^\epsilon}=1$ and, if $\epsilon$ is sufficiently small, $\ave{(p^\epsilon)^2+(q^\epsilon)^2}=1+(\kappa-2)\epsilon+o(\epsilon)$, hence $\ave{(p^\epsilon)^2+(q^\epsilon)^2}<1$ if $\kappa<2$. Therefore, owing to the discussion set forth in Section~\ref{sect:int.dyn}, in the regime of small $\epsilon$ the scaling~\eqref{eq:quasi-invariant_scaling} allows one to observe physical dynamics with conserved mean and internal energy evolving towards a finite non-zero value. Moreover, considering that the variation of the microscopic state due to the interaction is $v'-v=(p^\epsilon-1)v+q^\epsilon v_\ast$, we further observe that the parameter
$$ \kappa=\abs{\frac{\Var(p^\epsilon-1)}{\ave{p^\epsilon-1}}} $$
is the ratio between the stochastic and the deterministic average contributions of the $v$-coefficient $p^\epsilon$ to the post-interaction variation of the microscopic state $v$ itself. Conversely, since $\Var(q^\epsilon)=o(\Var(p^\epsilon))$ for $\epsilon\to 0^+$, the scaling~\eqref{eq:quasi-invariant_scaling} implies that the stochastic contribution of the $v_\ast$-coefficient $q^\epsilon$ to the variation of the microscopic state $v$ is negligible in the limit with respect to that of $p^\epsilon$.

As far as the transition probability is concerned, we scale $\beta$ as
$$ \beta^\epsilon(v,t)=\epsilon\beta(v,t), $$
so that from~\eqref{eq:trans_prob.death} we deduce $p(v,t;\,2\vert 1)\to 0$ and $p(v,t;\,1\vert 1)\to 1$ when $\epsilon\to 0^+$, meaning that also the label switching tends to be quasi-invariant (in probability) for $\epsilon$ small enough.

To compensate for the smallness of each interaction and each label switching, we simultaneously scale the corresponding rates as
\begin{equation}
	\lambda=\mu=\frac{1}{\epsilon},
	\label{eq:lambda_mu.scaling}
\end{equation}
which imply a high number of interactions and instances of label switch per unit time when $\epsilon\approx 0$.

Let us denote by $f_1^\epsilon(v,t)$ the distribution function of the group $i=1$ parametrised by the scaling parameter $\epsilon$. From~\eqref{eq:death_f1_weak} we deduce that it satisfies
\begin{align}
	\begin{aligned}[b]
		\frac{d}{dt}\int_{\R_+}\varphi(v)f_1^\epsilon(v,t)\,dv &= -\int_{\R_+}\varphi(v)\beta(v,t)f_1^\epsilon(v,t)\,dv \\
		&\phantom{=} +\frac{1}{\epsilon}\int_{\R_+}\int_{\R_+}\ave{\varphi(v')-\varphi(v)}f_1^\epsilon(v,t)f_1^\epsilon(v_\ast,t)\,dv\,dv_\ast.
	\end{aligned}
	\label{eq:q.i_inter}
\end{align}
Now, let $\varphi$ be a smooth and compactly supported function. Expanding the difference $\varphi(v')-\varphi(v)$ in Taylor series about $v$ and using~\eqref{eq:binary_gen_sym} with $p=p^\epsilon$, $q=q^\epsilon$ like in~\eqref{eq:quasi-invariant_scaling} we get
\begin{align}
	\begin{aligned}[b]
		\frac{d}{dt}\int_{\R_+}\varphi(v)f_1^\epsilon(v,t)\,dv &= -\int_{\R_+}\varphi(v)\beta(v,t)f_1^\epsilon(v,t)\,dv \\
		&\phantom{=} +\int_{\R_+}\int_{\R_+}\varphi'(v)(v_\ast-v)f_1^\epsilon(v,t)f_1^\epsilon(v_\ast,t)\,dv\,dv_\ast \\
		&\phantom{=} +\frac{\kappa\rho_1^\epsilon(t)}{2}\int_{\R_+}\varphi''(v)v^2f_1^\epsilon(v,t)\,dv+R_\varphi(f_1^\epsilon,\,f_1^\epsilon)(v,t),
	\end{aligned}
	\label{eq:FP+R}
\end{align}
where the remainder $R_\varphi(f_1^\epsilon,\,f_1^\epsilon)$ satisfies\footnote{Here and henceforth we use the notation $a\lesssim b$ to mean that there exists a constant $C>0$, independent of $\epsilon$ and whose specific value is unimportant, such that $a\leq Cb$.} (cf.~\cite{cordier2005JSP} for similar calculations)
\begin{align}
	\begin{aligned}[b]
		\abs{R_\varphi(f_1^\epsilon,\,f_1^\epsilon)(v,t)} &\lesssim \Vert\varphi''\Vert_\infty\left(\epsilon+\epsilon^\theta\right)\int_{\R_+}v^2f_1^\epsilon(v,t)\,dv \\
		&\phantom{\lesssim} +\Vert\varphi'''\Vert_\infty\left(\sqrt{\epsilon}+\epsilon^{\frac{1+3\theta}{2}}+\epsilon^2\right)\int_{\R_+}v^3f_1^\epsilon(v,t)\,dv.
	\end{aligned}
	\label{eq:remainder}
\end{align}

In view of the scaling~\eqref{eq:quasi-invariant_scaling}, we can standardise $p^\epsilon$, $q^\epsilon$ as
$$ p^\epsilon=1-\epsilon+\sqrt{\kappa\epsilon}Z, \qquad q^\epsilon=\epsilon+\sqrt{\kappa\epsilon^{1+\theta}}Z_\ast, $$
where $Z,\,Z_\ast$ are two independent random variables with zero mean and unitary variance, which we assume to be such that $\ave{\abs{Z}^3},\,\ave{\abs{Z_\ast}^3}<+\infty$. Thanks to this representation, setting $\varphi(v)=v^2,\,v^3$ in~\eqref{eq:q.i_inter} we further discover, after some algebraic calculations using in particular the inequalities $ab\leq\frac{1}{2}(a^2+b^2)$ and $ab^2\leq\frac{2}{3}(a^3+b^3)$ for $a,\,b\geq 0$, that
\begin{align*}
	& \frac{d}{dt}\int_{\R_+}v^2f_1^\epsilon(v,t)\,dv\lesssim\left(1+\epsilon+\epsilon^\theta\right)\int_{\R_+}v^2f_1^\epsilon(v,t)\,dv \\
	& \frac{d}{dt}\int_{\R_+}v^3f_1^\epsilon(v,t)\,dv\lesssim\left(1+\sqrt{\epsilon}+\epsilon+\epsilon^2+\epsilon^\theta+\epsilon^\frac{1+3\theta}{2}+\epsilon^{1+\theta}\right)
		\int_{\R_+}v^3f_1^\epsilon(v,t)\,dv,
\end{align*}
which imply that, for all fixed $t>0$, the terms $\int_{\R_+}v^2f_1^\epsilon(v,t)\,dv$, $\int_{\R_+}v^3f_1^\epsilon(v,t)\,dv$ remain bounded when $\epsilon\to 0^+$. Therefore, from~\eqref{eq:remainder} we infer
$$ R_\varphi(f_1^\epsilon,\,f_1^\epsilon)\xrightarrow{\epsilon\to 0^+}0. $$

Let us assume now that $(f_1^\epsilon)$ converges in $C(\R_+;\,L^1(\R_+)\cap L^1(\R_+;\,v\,dv))$, possibly up to subsequences, to a distribution function $f_1$ when $\epsilon\to 0^+$. Hence we have in particular
\begin{align*}
	& \rho_1^\epsilon(t)=\int_{\R_+}f_1^\epsilon(v,t)\,dv\xrightarrow{\epsilon\to 0^+}\rho_1(t)=\int_{\R_+}f_1(v,t)\,dv \\
	& \rho_1^\epsilon(t)m_1^\epsilon(t)=\int_{\R_+}vf_1^\epsilon(v,t)\,dv\xrightarrow{\epsilon\to 0^+}\rho_1(t)m_1(t)=\int_{\R_+}vf_1(v,t)\,dv,
\end{align*}
where $m_1^{(\epsilon)}(t):=\frac{1}{\rho_1^{(\epsilon)}(t)}\int_{\R_+}vf_1^{(\epsilon)}(v,t)\,dv$ denotes the first moment of the distribution function $f_1^{(\epsilon)}$. Then, passing to the limit $\epsilon\to 0^+$ in~\eqref{eq:FP+R} we obtain the limit equation
\begin{align*}
	\frac{d}{dt}\int_{\R_+}\varphi(v)f_1(v,t)\,dv &= -\int_{\R_+}\varphi(v)\beta(v,t)f_1(v,t)\,dv \\
	&\phantom{=} +\int_{\R_+}\int_{\R_+}\varphi'(v)(v_\ast-v)f_1(v,t)f_1(v_\ast,t)\,dv\,dv_\ast \\
	&\phantom{=} +\frac{\kappa\rho_1(t)}{2}\int_{\R_+}\varphi''(v)v^2f_1(v,t)\,dv,
\end{align*}
which, by integration by parts and recalling the compactness of the support of $\varphi$, can be recognised as a weak form of the following Fokker-Planck equation with non-constant coefficients and reaction term:
\begin{equation}
	\partial_tf_1=\frac{\kappa\rho_1(t)}{2}\partial_v^2(v^2f_1)+\rho_1(t)\partial_v\bigl((v-m_1(t))f_1\bigr)-\beta(v,t)f_1.
	\label{eq:FP.death.f1}
\end{equation}

The same quasi-invariant limit applied to~\eqref{eq:death_f2_weak} leads to
\begin{equation}
	\partial_tf_2=\beta(t,v)f_1.
	\label{eq:FP.death.f2}
\end{equation}

If we look for the stationary distributions, say $f_1^\infty$, $f_2^\infty$, in the quasi-invariant regime we find that they satisfy the system of equations
\begin{equation*}
	\begin{cases}
		\dfrac{\kappa\rho_1^\infty}{2}\partial_v^2(v^2f_1^\infty)+\rho_1^\infty\partial_v\bigl((v-m_1^\infty)f_1^\infty\bigr)-\beta^\infty(v)f_1^\infty=0 \\[2mm]
		\beta^\infty(v)f_1^\infty=0,
	\end{cases}
\end{equation*}
where the symbols $\rho_1^\infty$, $m_1^\infty$ have an obvious meaning while $\beta^\infty(v):=\lim_{t\to +\infty}\beta(v,t)$.

If $\beta^\infty$ is not identically zero then the second equation implies $f_1^\infty(v)=0$, which is clearly also a solution of the first equation. This is consistent with the idea that, in the long run, all living agents labelled with $i=1$ die by switching to the label $i=2$. Conversely, if $\beta^\infty\equiv 0$ then we distinguish two cases:
\begin{enumerate}[label=(\roman*)]
\item if $\beta(v,t)\to 0^+$ for $t\to +\infty$ but there exists $\beta_0=\beta_0(t)>0$ such that $\beta(v,t)\geq\beta_0(t)$ for all $t>0$ and all $v\in\R_+$ and moreover $\int_{\R_+}\beta_0(t)\,dt=+\infty$ (i.e., roughly speaking, $\beta(v,t)$ tends to zero slowly enough in time) then from~\eqref{eq:death.mass_balance} with the quasi-invariant scaling $\lambda=\frac{1}{\epsilon}$, $\beta_0^\epsilon(t)=\epsilon\beta_0(t)$ we deduce
$$ \rho_1(t)\leq\rho_{1,0}\exp{\left(-\int_0^t\beta_0(s)\,ds\right)}\xrightarrow{t\to +\infty}0, $$
whence we obtain again the stationary distribution $f_1^\infty(v)=0$;
\item if there exists $t_0>0$ such that $\beta(v,t)\equiv 0$ for $t\geq t_0$ then from $t_0$ onwards the masses $\rho_1$, $\rho_2$ are conserved. Moreover, owing to the quasi-invariant scaling~\eqref{eq:quasi-invariant_scaling}, also the first moment $m_1$ is conserved. Indeed from~\eqref{eq:q.i_inter} with $\varphi(v)=v$ we get, for $t>t_0$, $\frac{d}{dt}(\rho_1^\epsilon m_1^\epsilon)=0$ and the result follows passing to the limit $\epsilon\to 0^+$. Hence $\rho_1^\infty=\rho_1(t_0)>0$, $m_1^\infty=m_1(t_0)$ and the stationary distribution $f_1^\infty$ satisfies
$$ \frac{\kappa}{2}\partial_v(v^2f_1^\infty)+(v-m_1(t_0))f_1^\infty=0, $$
whose unique solution with mass $\rho_1(t_0)$ and first moment $m_1(t_0)$ is
$$ f_1^\infty(v)=\rho_1(t_0)\frac{(2\kappa m_1(t_0))^{2\kappa+1}}{\Gamma(2\kappa+1)}\cdot\frac{e^{-\frac{2\kappa m_1(t_0)}{v}}}{v^{2(\kappa+1)}}, $$
namely an inverse-gamma-type distribution with shape parameter $2\kappa+1$ and scale parameter $2\kappa m_1(t_0)$. Notice however that the exact determination of $\rho_1(t_0)$, $m_1(t_0)$ requires to solve the transient dynamics described by~\eqref{eq:FP.death.f1} up to $t=t_0$. The same is also necessary for the determination of $f_2^\infty(v)=f_2(v,t_0)$.
\end{enumerate}

\subsubsection{An explicitly solvable case}
Further insights into the solutions to~\eqref{eq:death_f1_weak},~\eqref{eq:death_f2_weak} can be obtained in the particular case in which the transition probability $\beta(v,t)$ is constant, say $\beta(v,t)\equiv\beta_0>0$. Then from~\eqref{eq:death_f1_weak} with $\varphi(v)=1,\,v$, together with the quasi-invariant scaling $\lambda=\frac{1}{\epsilon}$, $\beta_0^\epsilon=\epsilon\beta_0$ plus~\eqref{eq:quasi-invariant_scaling}, we obtain respectively, for $\epsilon\to 0^+$,
$$ \frac{d\rho_1}{dt}=-\beta_0\rho_1, \qquad \frac{d}{dt}(\rho_1m_1)=-\beta_0\rho_1m_1, $$
which imply $\rho_1(t)=\rho_{1,0}e^{-\beta_0t}$ and $m_1\equiv\text{constant}$, i.e. the first moment of $f_1$ is conserved in time. In this situation, it is reasonable to look for a self-similar solution of the form
$$ f_1(v,t)=\frac{\rho_1(t)}{m_1}g\!\left(\frac{v}{m_1}\right), $$
where $g:\R_+\to\R_+$ is such that
\begin{equation}
	\int_{\R_+}g(v)\,dv=1, \qquad \int_{\R_+}vg(v)\,dv=1.
	\label{eq:g.self-similar}
\end{equation}
Plugging into~\eqref{eq:FP.death.f1}, we discover that $g$ satisfies the following stationary Fokker-Planck equation:
\begin{equation}
	\frac{\kappa}{2}\partial_v^2(v^2g)+\partial_v\bigl((v-1)g\bigr)=0,
	\label{eq:FP.g.death}
\end{equation}
whose unique solution with unitary mass is
\begin{equation}
	g(v)=\frac{(2\kappa)^{2\kappa+1}}{\Gamma(2\kappa+1)}\cdot\frac{e^{-\frac{2\kappa}{v}}}{v^{2(\kappa+1)}},
	\label{eq:g}
\end{equation}
namely an inverse-gamma distribution with shape parameter $2\kappa+1$ and scale parameter $2\kappa$. Consequently, we determine
$$ f_1(v,t)=\rho_{1,0}e^{-\beta_0t}\frac{(2\kappa m_1)^{2\kappa+1}}{\Gamma(2\kappa+1)}\cdot\frac{e^{-\frac{2\kappa m_1}{v}}}{v^{2(\kappa+1)}} $$
and from~\eqref{eq:FP.death.f2}
$$ f_2(v,t)=f_{2,0}(v)+\rho_{1,0}\left(1-e^{-\beta_0t}\right)\frac{(2\kappa m_1)^{2\kappa+1}}{\Gamma(2\kappa+1)}\cdot\frac{e^{-\frac{2\kappa m_1}{v}}}{v^{2(\kappa+1)}}, $$
where $f_{2,0}(v)\geq 0$ with $\int_{\R_+}f_{2,0}(v)\,dv=1-\rho_{1,0}$ is the initial distribution function of the agents with label $i=2$. These solutions provide the exact evolution of the system under the joint label switch and interaction processes.

\subsection{Birth}
We consider now the birth process, in which the group $i=1$ composed of interacting agents accepts new incomes from the inert group $i=2$. The transition probabilities may therefore be chosen as
\begin{equation*}
	\begin{array}{ll}
		T(v,t;\,1\vert 2)=\beta(v,t), & T(v,t;\,2\vert 2)=1-\beta(v,t) \\
		T(v,t;\,2\vert 1)=0, & T(v,t;\,1\vert 1)=1
	\end{array}
\end{equation*}
with $0\leq\beta(v,t)\leq 1$ for all $t>0$ and all $v\in\R_+$. The kinetic equations describing the evolution of $f_1$ and $f_2$ can be deduced from~\eqref{eq:boltz.fi-same_label}, considering that the agents of the population $i=2$ do not interact. Therefore, we have
\begin{align}
	\begin{aligned}[b]
		\frac{d}{dt}\int_{\R_+}\varphi(v)f_1(v,t)\,dv &= \lambda\int_{\R_+}\varphi(v)\beta(v,t)f_2(v,t)\,dv \\
		&\phantom{=} +\mu\int_{\R_+}\int_{\R_+}\ave{\varphi(v')-\varphi(v)}f_1(v,t)f_1(v_\ast,t)\,dv\,dv_\ast
	\end{aligned}
	\label{eq:birth_f1_weak}
\end{align}
and
\begin{equation}
	\frac{d}{dt}\int_{\R_+}\varphi(v)f_2(v,t)\,dv=-\lambda\int_{\R_+}\varphi(v)\beta(v,t)f_2(v,t)\,dv.
	\label{eq:birth_f2_weak}
\end{equation}
Notice that in this case from~\eqref{eq:birth_f2_weak} we obtain explicitly
$$ f_2(v,t)=f_{2,0}(v)\exp\left(-\lambda\int_0^t \beta(v,s)\,ds\right), $$
which can be possibly plugged into~\eqref{eq:birth_f1_weak} to obtain a self-consistent equation for the sole distribution function $f_1$.

\subsubsection{Mass balance}
\label{sect:birth.mass_balance}
The evolution of the mass of the two populations is obtained with $\varphi(v)=1$ in~\eqref{eq:birth_f1_weak},~\eqref{eq:birth_f2_weak} and reads
\begin{align*}
	& \frac{d\rho_1}{dt}=\lambda\int_{\R_+}\beta(v,t)f_2(v,t)\,dv \\
	& \frac{d\rho_2}{dt}=-\lambda\int_{\R_+}\beta(v,t)f_2(v,t)\,dv.
\end{align*}

Like in Section~\ref{sect:death.mass}, if $\beta$ does not depend on $v$, i.e. $\beta=\beta(t)$, we determine explicitly
$$ \rho_2(t)=\rho_{2,0}\exp{\left(-\lambda\int_0^t\beta(s)\,ds\right)} $$
and consequently, recalling that $\rho_{1,0}+\rho_{2,0}=1$,
$$ \rho_1(t)=1-\rho_{2,0}\exp{\left(-\lambda\int_0^t\beta(s)\,ds\right)}. $$
In this case, if $\int_{\R_+}\beta(t)\,dt=+\infty$ then $\rho_2\to 0$ and $\rho_1\to 1$ for $t\to +\infty$, i.e. the whole population $i=2$ is born in the long run. If instead $\beta$ vanishes definitively from a certain time $t=t_0$ onwards then a residual mass of agents in $i=2$ remains and $\rho_1<1$ for $t\to +\infty$.

If $\beta$ features a full dependence on $t$ and $v$ then it is not possible to determine explicitly the evolution of $\rho_1$, $\rho_2$. Nevertheless, if there exists $\beta_0=\beta_0(t)\geq 0$ such that $\beta(v,t)\geq\beta_0(t)$ for all $t\geq 0$ and all $v\in\R_+$ then we may estimate
$$ \rho_1(t)\geq 1-\rho_{2,0}\exp{\left(-\lambda\int_0^t\beta_0(s)\,ds\right)}, \qquad \rho_2(t)\leq\rho_{2,0}\exp{\left(-\lambda\int_0^t\beta_0(s)\,ds\right)}, $$
which still imply $\rho_1\to 1$ and $\rho_2\to 0$ as $t\to +\infty$ if $\int_0^{+\infty}\beta_0(t)\,dt=+\infty$.

\subsubsection{Quasi-invariant limit and explicit solutions}
The same quasi-invariant scaling~\eqref{eq:quasi-invariant_scaling},~\eqref{eq:lambda_mu.scaling} of Section~\ref{sect:death.quasi-invariant_limit} applied to~\eqref{eq:birth_f1_weak},~\eqref{eq:birth_f2_weak} produces in this case
\begin{equation}
	\partial_tf_1=\frac{\kappa\rho_1(t)}{2}\partial_v^2(v^2f_1)+\rho_1(t)\partial_v\bigl((v-m_1(t))f_1\bigr)+\beta(v,t)f_2
	\label{eq:FP.birth.f1}
\end{equation}
and
\begin{equation}
	\partial_tf_2=-\beta(v,t)f_2.
	\label{eq:FP.birth.f2}
\end{equation}

From~\eqref{eq:FP.birth.f2}, we obtain that at the steady state it results $\beta^\infty(v)f_2^\infty=0$, hence from~\eqref{eq:FP.birth.f1} that $f_1^\infty$ solves
$$ \frac{\kappa}{2}\partial_v^2(v^2f_1^\infty)+\partial_v\bigl((v-m_1^\infty)f_1^\infty\bigr)=0. $$
Hence we deduce that $f_1^\infty$ is always an inverse-gamma-type distribution with mean $m_1^\infty$ and mass $\rho_1^\infty=\lim_{t\to+\infty}\rho_1(t)$. Using the arguments of Section~\ref{sect:birth.mass_balance}, we observe that $\rho_1^\infty=1$ whenever $\beta(v,t)$ tends to zero slowly enough for $t\to +\infty$. Otherwise, if $\beta(v,t)\equiv 0$ for $t\geq t_0$ then $\rho_1^\infty=\rho_1(t_0)\leq 1$.

Let us now consider~\eqref{eq:FP.birth.f1},~\eqref{eq:FP.birth.f2} in the case of constant $\beta$, say $\beta(v,t)=\beta_0>0$ for all $t\geq 0$ and all $v\in\R_+$. Then~\eqref{eq:FP.birth.f2} yields $f_2(v,t)=f_{2,0}(v)e^{-\beta_0t}$ and the first moment $m_2$ is conserved, say $m_2(t)=m$ for all $t\geq 0$. Next, from~\eqref{eq:birth_f1_weak} with $\varphi(v)=v$, under the scaling~\eqref{eq:quasi-invariant_scaling},~\eqref{eq:lambda_mu.scaling} and in the quasi-invariant limit $\epsilon\to 0^+$, we deduce
$$ \frac{dm_1}{dt}=\beta_0\frac{\rho_2}{\rho_1}(m-m_1), $$
which shows that $m_1(t)\to m$ for $t\to +\infty$ and, if $m_{1,0}=m$, that $m_1$ is in turn conserved and equals $m$ at all times. Therefore, it makes sense to look for a self-similar solution of~\eqref{eq:FP.birth.f1} of the form
$$ f_1(v,t)=\frac{\rho_1(t)}{m}g\!\left(\frac{v}{m}\right), $$
where $g:\R_+\to\R_+$ satisfies~\eqref{eq:g.self-similar}. Plugging into~\eqref{eq:FP.birth.f1} we get
\begin{equation}
	\frac{\kappa}{2}\partial_v^2(v^2g)+\partial_v((v-1)g)=\frac{\beta_0m}{\rho_1^2}\left(\frac{\rho_2}{m}g-f_2\right)
	\label{eq:FP.g.birth}
\end{equation}
whence, choosing the initial shape of $f_2$ as $f_{2,0}(v)=\frac{\rho_{2,0}}{m}g\!\left(\frac{v}{m}\right)$ so that $f_2(v,t)=\frac{\rho_2(t)}{m}g\!\left(\frac{v}{m}\right)$, we recover for $g$ the Fokker-Planck equation~\eqref{eq:FP.g.death}. This allows us to conclude that the time evolution of $f_1$, $f_2$ is given explicitly by
$$ f_1(v,t)=\left(1-\rho_{2,0}e^{-\beta_0 t}\right)\frac{(2\kappa m)^{2\kappa+1}}{\Gamma(2\kappa+1)}\cdot\frac{e^{-\frac{2\kappa m}{v}}}{v^{2(\kappa+1)}}, \qquad 
	f_2(v,t)=\rho_{2,0}e^{-\beta_0 t}\frac{(2\kappa m)^{2\kappa+1}}{\Gamma(2\kappa+1)}\cdot\frac{e^{-\frac{2\kappa m}{v}}}{v^{2(\kappa+1)}}. $$

If at the initial time $f_2$ is not the $g$-shaped distribution above then~\eqref{eq:FP.g.birth} cannot be solved explicitly due to the time-varying reaction term on the right-hand side. Nevertheless, since $f_2(v,t)\to 0$, $\rho_1(t)\to 1$ and $\rho_2(t)\to 0$ when $t\to +\infty$, we formally infer that for large times $g$ still solves~\eqref{eq:FP.g.death}. Consequently, we at least characterise the stationary distributions as
$$ f_1^\infty(v)=\frac{(2\kappa m)^{2\kappa+1}}{\Gamma(2\kappa+1)}\cdot\frac{e^{-\frac{2\kappa m}{v}}}{v^{2(\kappa+1)}}, \qquad f_2^\infty(v)=0. $$

\section{A kinetic model of the contagion of infectious diseases with quarantine}
\label{sect:contagion}
Models for the spreading of infectious diseases are a prominent example of dynamics in which agents switch from one group to another depending on their infection condition. Classical population dynamics models take a super-macroscopic point of view and describe the time evolution of the total number of susceptible ($S$), infected ($I$) and recovered ($R$) individuals assuming that the probability of a contagious encounter between a susceptible and an infected individual is proportional to the size of either group.

As an example, we show that the popular SIR model may be derived from the kinetic description of the pure Markov-type jump process presented in Section~\ref{sect:label.switch}. Let $x=1$ label the susceptible individuals, $x=2$ the infected individuals and $x=3$ the recovered individuals. The kinetic distribution function $f$ may be represented by setting $n=3$ in~\eqref{eq:f.delta.label_switch}, $f_1$, $f_2$, $f_3$ being the probabilities (normalised masses) of the three groups of individuals at time $t$. They satisfy the system of equations~\eqref{eq:fi.jump} in which, up to a time scaling, we may conveniently assume $\lambda=1$. If we further specify the transition probabilities as
\begin{equation*}
	\begin{array}{lll}
		T(t;\,1\vert 1)=1-\beta f_2(t), & T(t;\,2\vert 1)=\beta f_2(t), & T(t;\,3\vert 1)=0 \\
		T(t;\,1\vert 2)=0, & T(t;\,2\vert 2)=1-\gamma, & T(t;\,3\vert 2)=\gamma \\
		T(t;\,1\vert 3)=0, & T(t;\,2\vert 3)=0, & T(t;\,3\vert 3)=1 \\
	\end{array}
\end{equation*}
with $0\leq\beta,\,\gamma\leq 1$, we obtain from~\eqref{eq:fi.jump}
$$	\begin{cases}
		f_1'=-\beta f_1f_2 \\
		f_2'=\beta f_1f_2-\gamma f_2 \\
		f_3'=\gamma f_2,
	\end{cases} $$
which has indeed the form of the SIR model. Notice that all the transition probabilities are constant but those associated with the label switch $1\to 2$ from susceptible to infected, which are proportional to the density of infected individuals.

As this example demonstrates, this type of models does not describe the progression of the contagion as the result of actual microscopic contacts among the individuals but simply as a consequence of their jumps from one label to another. The kinetic framework introduced in the previous sections allows us to conceive a richer model, in which individual contacts are taken into account in more detail. Consequently, me may describe contagion dynamics by invoking explicitly some relevant microscopic determinants, such as the viral load of the individuals. Our model, which shares formal analogies with the one presented in~\cite{delitala2004MCM}, is actually an \textit{epidemiological caricature} susceptible of several improvements towards a more accurate description of real-world scenarios. It has however the merit of introducing a microscopic individual-based characterisation of the disease, normally lacking in classical epidemiological models, while allowing us to address quite precisely a few interesting properties of the solutions, including the determination of hydrodynamic equations and of the asymptotic equilibrium distributions.

In this application, we assume that the microscopic state $v\in\R_+$ of the individuals represents their \textit{viral load} and moreover that there are two groups of people: those labelled by $x=1$, who have not been diagnosed with the infection yet, and those labelled by $x=2$, who have been detected as infected and quarantined. Undiagnosed people ($x=1$) interact with one another and switch to label $x=2$ once diagnosed. Conversely, quarantined people ($x=2$) do not interact due to isolation and possibly switch back to label $x=1$ when their viral load has decreased enough. To describe these dynamics, the reference framework is provided by the kinetic equation~\eqref{eq:boltz.fi-same_label}, which takes into account interactions within the same labels only. In particular, we model the transition probabilities as
\begin{subequations}
	\begin{alignat}{2}
		& T(v;\,1\vert 1)=1-\alpha(v), && \qquad T(v;\,2\vert 1)=\alpha(v) \label{eq:T_quarantine.1->2} \\
		& T(v;\,1\vert 2)=\beta(v), && \qquad T(v;\,2\vert 2)=1-\beta(v), \label{eq:T_quarantine.2->1}
	\end{alignat}
	\label{eq:T_quarantine}
\end{subequations}
where $0\leq\alpha(v),\,\beta(v)\leq 1$ are the probabilities that an individual with viral load $v$ is diagnosed and quarantined or is readmitted in the society, respectively. We assume that these probabilities are time-independent and moreover that $\alpha$ is non-decreasing and $\beta$ is non-increasing in $v$.

As far as the interaction rules are concerned, we assume that undiagnosed people ($x=1$) may infect other undiagnosed people depending on their current viral load. Specifically, we set
\begin{equation}
	v'=(1-\nu_1+\eta)v+\nu_2v_\ast
	\label{eq:int.undiagnosed}
\end{equation}
where $\nu_1,\,\nu_2\in (0,\,1)$ are exchange rates among the individuals modelling the contagion dynamics and $\eta\in (\nu_1-1,\,+\infty)$ is a centred (i.e. $\ave{\eta}=0$) random variable accounting for random fluctuations in the individual viral load. Notice that~\eqref{eq:int.undiagnosed} is of the form~\eqref{eq:binary_gen_sym} with stochastic $p=1-\nu_1+\eta$ and deterministic $q=\nu_2$.

Conversely, we assume that quarantined people ($x=2$) may only recover from the infection due to the lack of contacts with other individuals:
\begin{equation}
	v'=(1-\gamma+\xi)v,
	\label{eq:int.quarantined}
\end{equation}
where $\gamma\in (0,\,1)$ is the rate of recovery and $\xi\in (\gamma-1,\,+\infty)$ is another centred random variable independent of $\eta$. Also~\eqref{eq:int.quarantined} is of the form~\eqref{eq:binary_gen_sym} with stochastic $p=1-\gamma+\xi$ and deterministic $q=0$.

Consequently, the equations for the distribution functions $f_1$, $f_2$ of the undiagnosed and quarantined people read:
\begin{align}
	\begin{aligned}[b]
		\frac{d}{dt}\int_{\R_+}\varphi(v)f_1(v,t)\,dv &= \lambda\int_{\R_+}\varphi(v)\Bigl(\beta(v)f_2(v,t)-\alpha(v)f_1(v,t)\Bigr)dv \\
		&\phantom{=} +\mu\int_{\R_+}\int_{\R_+}\ave{\varphi(v')-\varphi(v)}f_1(v,t)f_1(v_\ast,t)\,dv\,dv_\ast
	\end{aligned}
	\label{eq:undiagnosed}
\end{align}
and
\begin{align}
	\begin{aligned}[b]
		\frac{d}{dt}\int_{\R_+}\varphi(v)f_2(v,t)\,dv &= \lambda\int_{\R_+}\varphi(v)\Bigl(\alpha(v)f_1(v,t)-\beta(v)f_2(v,t)\Bigr)dv \\
		&\phantom{=} +\mu\int_{\R_+}\ave{\varphi(v')-\varphi(v)}f_2(v,t)\,dv.
	\end{aligned}
	\label{eq:quarantined}
\end{align}
We observe that, unlike the general equation~\eqref{eq:boltz.fi}, in~\eqref{eq:quarantined} the interaction term (second term on the right-hand side) is linear with respect to $f_2$. The reason is that~\eqref{eq:int.quarantined} is not a binary interaction but simply an update of the state of a quarantined individual based on their current state only. Equation~\eqref{eq:quarantined} may be obtained from~\eqref{eq:boltz.fi} by assuming that $\bar{v}'=\tilde{v}'$ are independent of $v_\ast$ and using~\eqref{eq:f.prob}.

\begin{remark}
Recently, another kinetic model dealing with the spreading of infectious diseases and which may be seen as a particular case of the general framework presented in this paper has been proposed~\cite{dimarco2020PRE}. However, contagion dynamics are not the main focus of that work, indeed they are described by a kinetic rephrasing of the SIR model analogous to the one discussed ad the beginning of this section. The focus is instead on the impact of infectious diseases on the socio-economic status of the individuals.
\end{remark}

\subsection{Constant transition probabilities}
\label{sect:alpha_beta.const}
We investigate first the simple case of constant $\alpha,\,\beta$, which allows for a detailed analysis of the trend of model~\eqref{eq:undiagnosed}-\eqref{eq:quarantined}.

Assume
$$ \lambda=\mu=1, $$
meaning that the label switch and the interactions take place on the same time scale. Setting $\varphi(v)=1,\,v$ in~\eqref{eq:undiagnosed},~\eqref{eq:quarantined} and using the interaction rules~\eqref{eq:int.undiagnosed},~\eqref{eq:int.quarantined}, we obtain the following equations for the evolution of the zeroth and first moments of $f_1,\,f_2$:
\begin{subequations}
	\begin{empheq}[left=\empheqlbrace]{align}
		& \frac{d\rho_1}{dt}=-\alpha\rho_1+\beta\rho_2 \label{eq:rho1} \\
		& \frac{d\rho_2}{dt}=\alpha\rho_1-\beta\rho_2 \label{eq:rho2} \\
		& \frac{d}{dt}(\rho_1m_1)=-[\alpha+(\nu_1-\nu_2)\rho_1]\rho_1m_1+\beta\rho_2m_2 \label{eq:rho1m1} \\
		& \frac{d}{dt}(\rho_2m_2)=\alpha\rho_1m_1-(\beta+\gamma)\rho_2m_2. \label{eq:rho2m2}
	\end{empheq}
\end{subequations}
From~\eqref{eq:rho1},~\eqref{eq:rho2}, together with the natural initial conditions $\rho_{1,0}=1$, $\rho_{2,0}=0$, we obtain
\begin{equation}
	\rho_1(t)=\frac{\beta}{\alpha+\beta}\left(1+\frac{\alpha}{\beta}e^{-(\alpha+\beta)t}\right), \qquad
		\rho_2(t)=\frac{\alpha}{\alpha+\beta}\left(1-e^{-(\alpha+\beta)t}\right),
	\label{eq:Tconst.rho}
\end{equation}
whence $\rho_1^\infty=\frac{\beta}{\alpha+\beta}$ and $\rho_2^\infty=\frac{\alpha}{\alpha+\beta}$. Therefore, quarantine apparently settles as a persistent condition involving a fixed fraction $\rho_2^\infty$ of the population. This may indicate that the infection cannot be eradicated and becomes endemic.

Nevertheless, we obtain a clearer picture by considering the further piece of information provided by~\eqref{eq:rho1m1},~\eqref{eq:rho2m2}, which unveil the evolution of the mean viral loads $m_1,\,m_2$ of undiagnosed and quarantined individuals. Using~\eqref{eq:rho1},~\eqref{eq:rho2}, we rewrite system~\eqref{eq:rho1m1}-\eqref{eq:rho2m2} in vector form as
\begin{equation*}
	\frac{d}{dt}
		\begin{pmatrix}
			m_1 \\
			m_2
		\end{pmatrix}
	=	\begin{pmatrix}
			(\nu_2-\nu_1)\rho_1-\beta\frac{\rho_2}{\rho_1} & \beta\frac{\rho_2}{\rho_1} \\
			\alpha\frac{\rho_1}{\rho_2} & -\left(\alpha\frac{\rho_1}{\rho_2}+\gamma\right)
		\end{pmatrix}
		\begin{pmatrix}
			m_1 \\
			m_2
		\end{pmatrix}.
\end{equation*}
Due to the presence of $\rho_1$, $\rho_2$ the system matrix is time-dependent, namely the system is non-autonomous. To approach the qualitative study of the large time trend of its trajectories we linearise the system around the equilibria $\rho_1=\rho_1^\infty$, $\rho_2=\rho_2^\infty$ and $m_1=m_2=0$. The goal is to investigate the stability and attractiveness of the asymptotic state expressing eradication of the infection, i.e.
$$ f_1^\infty(v)=\frac{\beta}{\alpha+\beta}\delta(v), \qquad f_2^\infty(v)=\frac{\alpha}{\alpha+\beta}\delta(v). $$
We obtain:
\begin{equation}
	\frac{d}{dt}
		\begin{pmatrix}
			m_1 \\
			m_2
		\end{pmatrix}
	=	\begin{pmatrix}
			(\nu_2-\nu_1)\frac{\beta}{\alpha+\beta}-\alpha & \alpha \\
			\beta & -(\beta+\gamma)
		\end{pmatrix}
		\begin{pmatrix}
			m_1 \\
			m_2
		\end{pmatrix},
	\label{eq:lin_sys.m1m2}
\end{equation}
that we study in two representative cases.

As a first example, let us consider $\nu_1=\nu_2$ in~\eqref{eq:int.undiagnosed}. With classical arguments of linear stability we find that the equilibrium $(m_1^\infty,\,m_2^\infty)=(0,\,0)$ is globally asymptotically stable if $\alpha>0$. Thus any arbitrarily small quarantine rate leads, in the long run, to the eradication of the infection. However, the smallness of $\alpha$ affects considerably the speed of convergence to such an equilibrium. Indeed, for $\alpha\to 0^+$ the eigenvalues $\omega_1,\,\omega_2$ of the linear system~\eqref{eq:lin_sys.m1m2} can be approximated as
$$ \omega_1=-\frac{\alpha\gamma}{\beta+\gamma}+o(\alpha), \qquad \omega_2=-(\beta+\gamma)+o(1). $$
The expression of $\omega_1$ shows that the convergence may be particularly slow.

As a second example, let us consider $\nu_1=0$, $\nu_2>0$ in~\eqref{eq:int.undiagnosed}. Hence an individual can only get more infected by the contact with another infected individual. Notice that without label switching, i.e. for $\alpha=\beta=0$, such a microscopic interaction would lead to a blow up of the mean viral load $m_1$ in time, indeed $\ave{p+q}=1+\nu_2>1$ (cf. Section~\ref{sect:int.dyn}). Conversely, thanks to the label switching, $m_1$ converges to zero if
$$ \alpha>\alpha_\dagger:=\max\left\{\max\left\{0,\,\frac{-(2\beta+\gamma)+\sqrt{\gamma^2+4\nu_2\beta}}{2}\right\},\,
	\frac{-\beta+\sqrt{\left(1+\frac{4\nu_2}{\gamma}\right)\beta^2+4\nu_2\beta}}{2}\right\}. $$
Clearly, the higher the contagion rate $\nu_2$ the more promptly infected individuals have to be diagnosed and quarantined.

\subsection{Variable transition probabilities: Two-scale dynamics and hydrodynamic limit}
\label{sect:alpha_beta.var}
For variable $\alpha$, $\beta$, a regime which allows us to gain some insights into the trends of model~\eqref{eq:undiagnosed}-\eqref{eq:quarantined} is when the label switching and the interaction processes take place on two well separated time scales. Assume
$$ \lambda=1, \qquad \mu=\frac{1}{\delta}, $$
where $0<\delta\ll 1$ is a small parameter. This implies that interactions among the agents with the same label are much more frequent than changes of label. In view of this argument, we can split e.g.,~\eqref{eq:undiagnosed} as
\begin{align*}
	&\frac{d}{dt}\int_{\R_+}\varphi(v)f_1(v,t)\,dv=\frac{1}{\delta}\int_{\R_+}\int_{\R_+}\ave{\varphi(v')-\varphi(v)}f_1(v,t)f_1(v_\ast,t)\,dv\,dv_\ast \\
	&\frac{d}{dt}\int_{\R_+}\varphi(v)f_1(v,t)\,dv=\int_{\R_+}\varphi(v)\Bigl(\beta(v)f_2(v,t)-\alpha(v)f_1(v,t)\Bigr)dv.
\end{align*}
By introducing the new time scale
$$ \tau:=\frac{t}{\delta} $$
and defining $\tilde{f}_1(v,\tau):=f_1(v,t)$, we rewrite the system above as
\begin{subequations}
	\begin{empheq}{align}
		&\frac{d}{d\tau}\int_{\R_+}\varphi(v)\tilde{f}_1(v,\tau)\,dv=\int_{\R_+}\int_{\R_+}\ave{\varphi(v')-\varphi(v)}\tilde{f}_1(v,\tau)\tilde{f}_1(v_\ast,\tau)\,dv\,dv_\ast 
			\label{eq:split.interaction.f1} \\
		&\frac{d}{dt}\int_{\R_+}\varphi(v)f_1(v,t)\,dv=\int_{\R_+}\varphi(v)\Bigl(\beta(v)f_2(v,t)-\alpha(v)f_1(v,t)\Bigr)dv, \label{eq:split.switching.f1}
	\end{empheq}
\end{subequations}
that we interpret as follows: while the conservative interaction dynamics~\eqref{eq:split.interaction.f1} reach rapidly an equilibrium on the quick time scale $\tau$, the label switching dynamics~\eqref{eq:split.switching.f1} are basically frozen on the slow time scale $t$.

The equilibrium on the $\tau$-scale is especially interesting when $\nu_1=\nu_2$ in~\eqref{eq:int.undiagnosed}, for then the contagion dynamics do not only conserve the mass $\rho_1$ of the undiagnosed individuals but also their mean viral load $m_1$. Indeed, in this case we have $\ave{p+q}=1$, cf. Section~\ref{sect:int.dyn}. As a consequence, the $\tau$-asymptotic distribution produced by~\eqref{eq:split.interaction.f1} is parametrised by both these macroscopic quantities on the $t$-scale and can be expressed in self-similar form as
$$ \frac{\rho_1(t)}{m_1(t)}g_1\!\left(\frac{v}{m_1(t)}\right), $$
where $g_1:\R_+\to\R_+$ satisfies the normalisation conditions
$$ \int_{\R_+}g_1(v)\,dv=1, \qquad \int_{\R_+}vg_1(v)\,dv=1. $$

An analogous splitting argument applied to~\eqref{eq:quarantined} produces
\begin{subequations}
	\begin{empheq}{align}
		&\frac{d}{d\tau}\int_{\R_+}\varphi(v)\tilde{f}_2(v,\tau)\,dv=\int_{\R_+}\ave{\varphi(v')-\varphi(v)}\tilde{f}_2(v,\tau)\,dv
			\label{eq:split.interaction.f2} \\
		&\frac{d}{dt}\int_{\R_+}\varphi(v)f_2(v,t)\,dv=\int_{\R_+}\varphi(v)\Bigl(\alpha(v)f_1(v,t)-\beta(v)f_2(v,t)\Bigr)dv \label{eq:split.switching.f2}
	\end{empheq}
\end{subequations}
with $\tilde{f}_2(v,\tau):=f_2(v,t)$. We now let $\rho_2(t)g_2(v)$ denote the $\tau$-asymptotic distribution generated by~\eqref{eq:split.interaction.f2}, where $g_2:\R_+\to\R_+$ satisfies only the normalisation condition
$$ \int_{\R_+}g_2(v)\,dv=1 $$
because the recovery dynamics~\eqref{eq:int.quarantined} conserve only the mass of the quarantined individuals.

On the whole, on the $t$-scale we express
\begin{equation}
	f_1(v,t)=\frac{\rho_1(t)}{m_1(t)}g_1\!\left(\frac{v}{m_1(t)}\right), \qquad f_2(v,t)=\rho_2(t)g_2(v)
	\label{eq:split.f1.f2}
\end{equation}
so that, plugging these distributions into~\eqref{eq:split.switching.f1},~\eqref{eq:split.switching.f2} and taking the conservation relationship $\rho_1(t)+\rho_2(t)=1$ into account, we obtain the evolution of the macroscopic parameters $\rho_1$, $\rho_2$, $m_1$ on the slow time scale $t$ analogously to what happens in classical kinetic theory with the \textit{hydrodynamic limit}:
\begin{equation}
	\begin{cases}
		\dfrac{d\rho_1}{dt}=\left(\displaystyle{\int_{\R_+}}\beta(v)g_2(v)\,dv\right)\rho_2-\left(\displaystyle{\int_{\R_+}}\alpha(m_1v)g_1(v)\,dv\right)\rho_1 \\
		\rho_2=1-\rho_1 \\
		\dfrac{d}{dt}(\rho_1m_1)=\left(\displaystyle{\int_{\R_+}}v\beta(v)g_2(v)\,dv\right)\rho_2-\left(\displaystyle{\int_{\R_+}}v\alpha(m_1v)g_1(v)\,dv\right)\rho_1m_1.
	\end{cases}
	\label{eq:split.hydro}
\end{equation}

It is not difficult to check that~\eqref{eq:split.interaction.f2}, together with the microscopic dynamics~\eqref{eq:int.quarantined}, produces $g_2(v)=\delta(v)$. Indeed,~\eqref{eq:int.quarantined} is such that $\ave{p+q}=1-\gamma<1$, which according to Section~\ref{sect:int.dyn} implies $m_2\to 0^+$ asymptotically in time. Therefore,~\eqref{eq:split.hydro} specialises as
\begin{equation*}
	\begin{cases}
		\dfrac{d\rho_1}{dt}=\beta(0)(1-\rho_1)-\left(\displaystyle{\int_{\R_+}}\alpha(m_1v)g_1(v)\,dv\right)\rho_1 \\
		\dfrac{d}{dt}(\rho_1m_1)=-\left(\displaystyle{\int_{\R_+}}v\alpha(m_1v)g_1(v)\,dv\right)\rho_1m_1,
	\end{cases}
\end{equation*}
whose steady states are given by
$$ \beta(0)-\left(\beta(0)+\displaystyle{\int_{\R_+}}\alpha(m_1^\infty v)g_1(v)\,dv\right)\rho_1^\infty=0, \quad
	\left(\displaystyle{\int_{\R_+}}v\alpha(m_1^\infty v)g_1(v)\,dv\right)\rho_1^\infty m_1^\infty=0. $$
	
Assume $\beta(0)>0$, then from the first equation we deduce $\rho_1^\infty>0$. Assume also that the mapping $v\mapsto\alpha(v)$ is strictly increasing with $\alpha(0)=0$. Then from the second equation we get $m_1^\infty=0$, considering that the integral term does not vanish if $m_1^\infty>0$. Indeed, from the assumed monotonicity of $\alpha$ we have
$$ \int_{\R_+}v\alpha(m_1^\infty v)g_1(v)\,dv\geq\alpha\!\left(\frac{m_1^\infty}{2}\right)\int_{\frac{1}{2}}^{+\infty}vg_1(v)\,dv $$
and moreover, from the normalisation properties of $g_1$,
$$ \int_{\frac{1}{2}}^{+\infty}vg_1(v)\,dv=\int_{\R_+}vg_1(v)\,dv-\int_0^{\frac{1}{2}}vg_1(v)\,dv\geq 1-\frac{1}{2}\int_0^{\frac{1}{2}}g_1(v)\,dv\geq\frac{1}{2}. $$

In conclusion, we obtain $\rho_1^\infty=1$ and $m_1^\infty=0$, which from~\eqref{eq:split.f1.f2} produce
$$ f_1^\infty(v)=\delta(v), \qquad f_2^\infty(v)=0. $$
Hence the quarantine leads, in the long run, to a full recovery of the population $x=1$ with no quarantined individuals.

\section{Numerical tests}
\label{sect:numerical}
We show now some numerical solutions to model~\eqref{eq:undiagnosed}-\eqref{eq:quarantined} with interaction rules~\eqref{eq:int.undiagnosed},~\eqref{eq:int.quarantined}, which confirm the findings of Sections~\ref{sect:alpha_beta.const},~\ref{sect:alpha_beta.var} and  allow us to explore also cases not explicitly covered by the previous theoretical study.

To solve the kinetic equations~\eqref{eq:undiagnosed}-\eqref{eq:quarantined} numerically we use a modified version of the Nanbu-Babovski Monte Carlo algorithm~\cite{bobylev2000PRE,pareschi2001ESAIMP,pareschi2013BOOK}, see Algorithm~\ref{alg:nanbu} in Appendix~\ref{app:nanbu}, which includes the transfer of agents from one label to another. Algorithm~\ref{alg:nanbu} is based on a direct implementation of the time discrete stochastic microscopic processes~\eqref{eq:micro.rules.gen}, which in the limit $\Delta{t}\to 0^+$ produce the kinetic equations. In particular, $\Theta,\,\Xi$ are distributed according to~\eqref{eq:bernoulli}, $J_t$ is conditionally distributed according to~\eqref{eq:T_quarantine} and $V_t'$ is defined like either~\eqref{eq:int.undiagnosed} or~\eqref{eq:int.quarantined} depending on the population label. The algorithm consists of two main blocks, which may be executed in parallel:
\begin{enumerate*}[label=\roman*)]
\item lines~\ref{algo:label_switching-start}--\ref{algo:label_switching-end} implement the label switch process;
\item lines~\ref{algo:interactions-start}--\ref{algo:interactions-end} implement the microscopic interactions.
\end{enumerate*}

In Table~\ref{tab:constant} we list the parameters of the algorithm and of the model that we keep fixed in all numerical tests. In Table~\ref{tab:variable} we list instead those that we vary from test to test. In all numerical tests we prescribe as initial conditions:
$$ f_{1,0}(v)=\mathbb{1}_{[0,\,1]}(v), \qquad f_{2,0}(v)=0. $$
Hence, $f_1^0$ is the uniform distribution in $[0,\,1]$ (from which we sample initially the particles with label $x=1$ in Algorithm~\ref{alg:nanbu}) while no agents are quarantined at $t=0$.

\begin{table}[!t]
\centering
\caption{Parameters kept constant in all numerical tests of Section~\ref{sect:numerical}}
\label{tab:constant}
\begin{tabular}{l|cccccc}
Parameter & $N$ & $\lambda$ & $\Delta{t}$ & $\nu_2$ & $\gamma$ \\
\hline
\hline
Value & $10^6$ & $1$ & $10^{-3}$ & $0.2$ & $0.3$ \\
\hline
\end{tabular}
\end{table}
\begin{table}[!t]
\centering
\caption{Parameters varying from test to test of Section~\ref{sect:numerical}}
\label{tab:variable}
\begin{tabular}{c|cccccc}
Parameter & Figure~\ref{fig:Tconst_stable} & Figure~\ref{fig:Tconst_unstable} & Figure~\ref{fig:Tvar_nu1=nu2_mu=10} & Figure~\ref{fig:Tvar_nu1=nu2_mu=1} & Figure~\ref{fig:Tvar_nu1=0_mu=10} & Figure~\ref{fig:Tvar_nu1=0_mu=1} \\
\hline
\hline
$\mu$ & $1$ & $1$ & $10$ & $1$ & $10$ & $1$ \\
$\alpha$ & $0.8$ & $0.2$ & $0.8\left(1-e^{-v}\right)$ & $0.8\left(1-e^{-v}\right)$ & $0.8\left(1-e^{-v}\right)$ & $0.8\left(1-e^{-v}\right)$ \\
$\beta$ & $0.4$ & $0.4$ & $0.4 e^{-v}$ & $0.4 e^{-v}$ & $0.4 e^{-v}$ & $0.4 e^{-v}$ \\
$\nu_1$ & $0$ & $0$ & $0.2$ & $0.2$ & $0$ & $0$ \\
$\alpha_\dagger$ & $0.28$ & $0.28$ & -- & -- & -- & -- \\
\hline
\end{tabular}
\end{table}

\begin{figure}[!t]
\centering
\subfigure[]{\includegraphics[width=.4\textwidth]{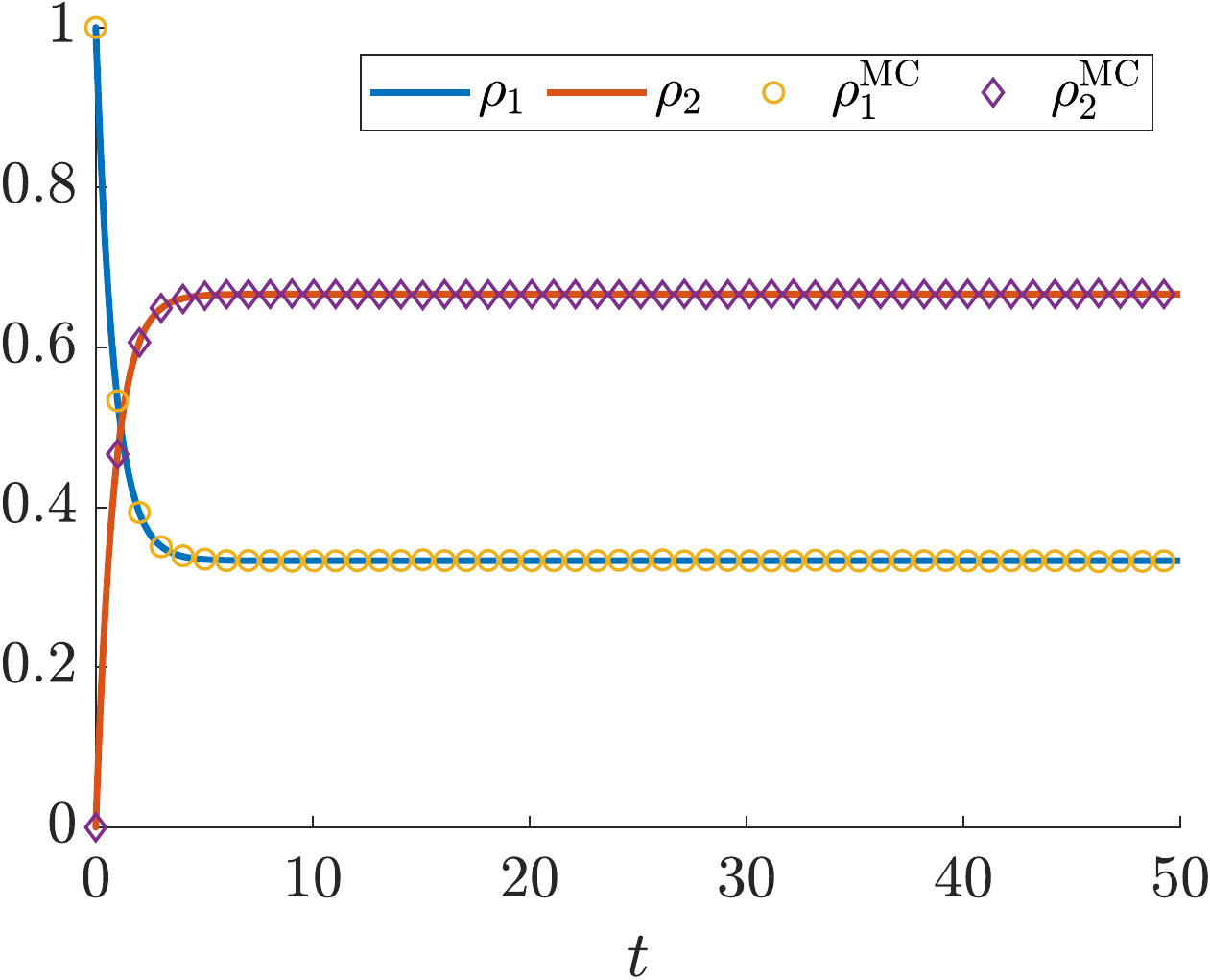}} \qquad
\subfigure[]{\includegraphics[width=.4\textwidth]{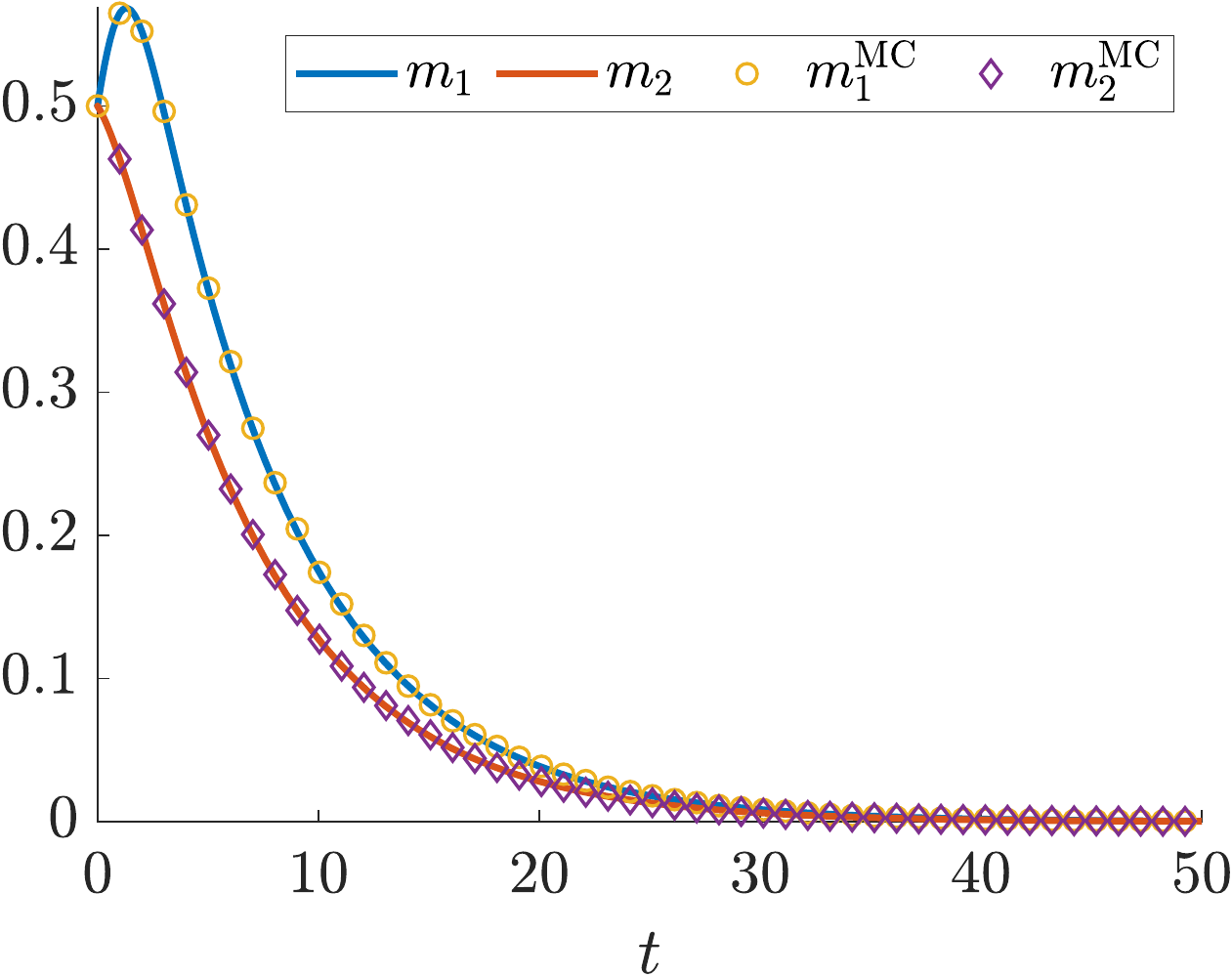}}
\caption{The problem of Section~\ref{sect:alpha_beta.const} with $\nu_1=0$ and $\alpha>\alpha_\dagger$. The continuous lines are the solutions of the hydrodynamic model~\eqref{eq:rho1}--\eqref{eq:rho2m2}. In particular, the solutions to~\eqref{eq:rho1}-\eqref{eq:rho2} are given exactly by~\eqref{eq:Tconst.rho} while the solutions to~\eqref{eq:rho1m1}-\eqref{eq:rho2m2} have been obtained numerically by a fourth order Runge-Kutta scheme}
\label{fig:Tconst_stable}
\end{figure}

\begin{figure}[!t]
\centering
\subfigure[]{\includegraphics[width=.4\textwidth]{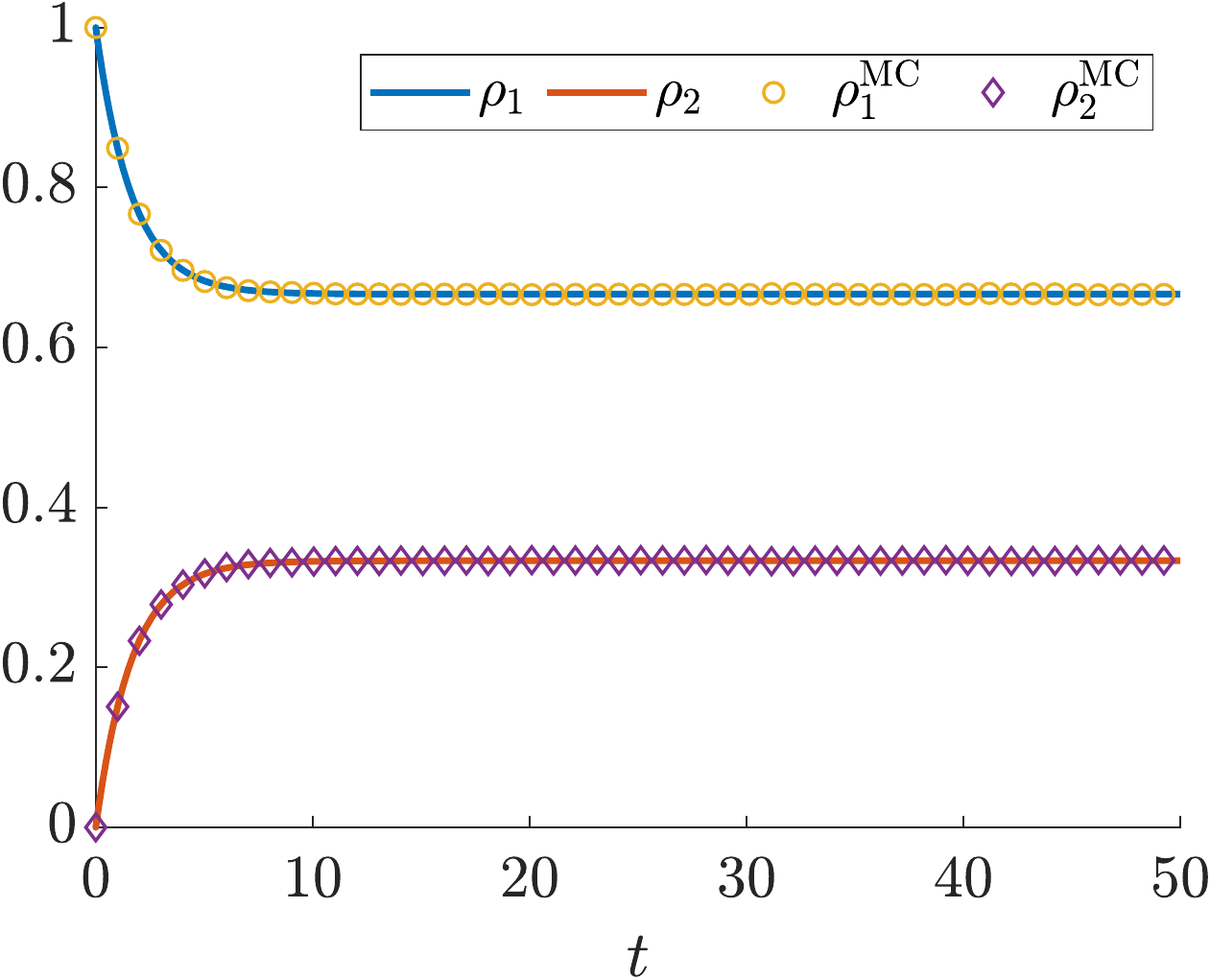}} \qquad
\subfigure[]{\includegraphics[width=.4\textwidth]{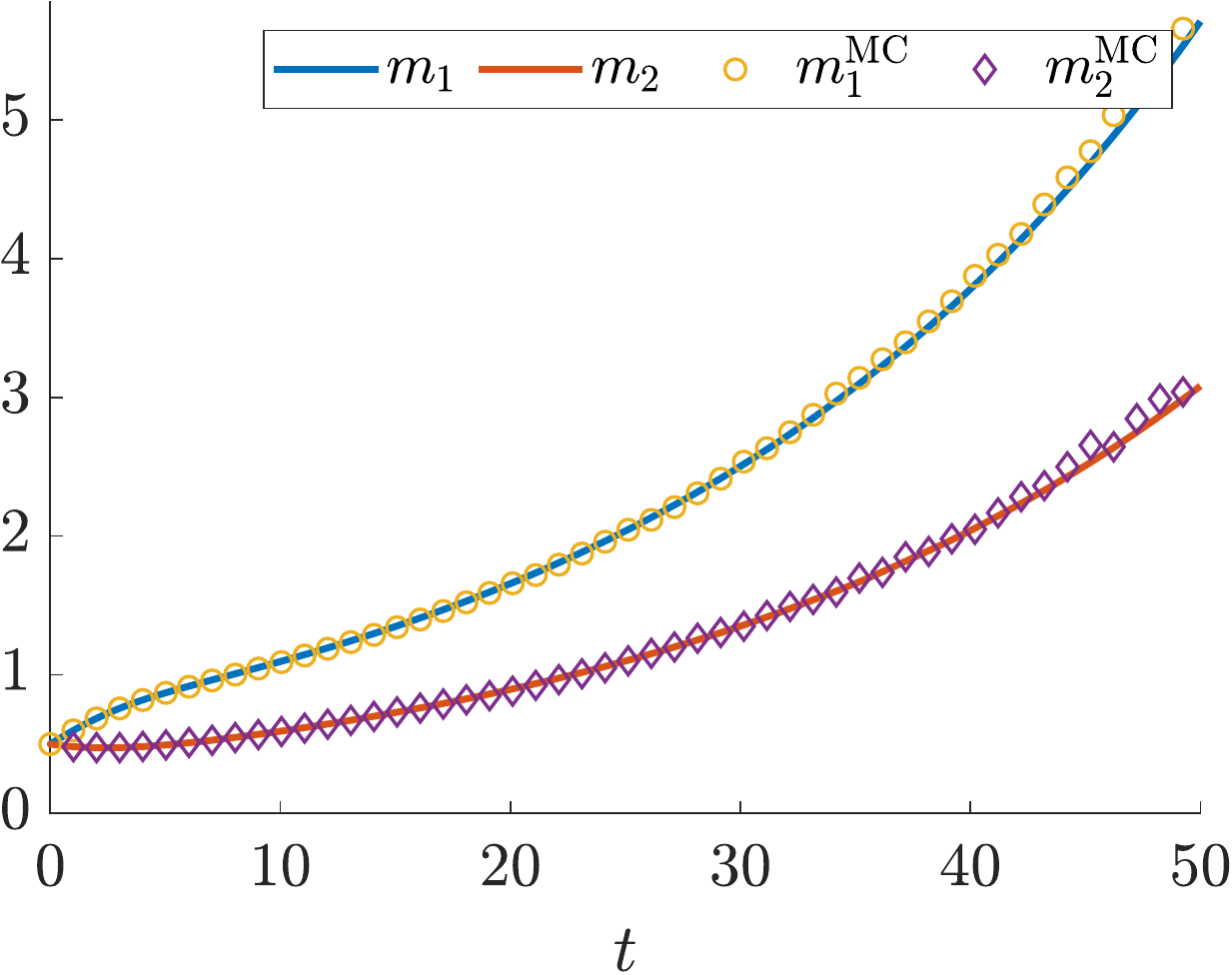}}
\caption{The problem of Section~\ref{sect:alpha_beta.const} with $\nu_1=0$ and $\alpha<\alpha_\dagger$. The continuous lines are the solutions of the hydrodynamic model~\eqref{eq:rho1}--\eqref{eq:rho2m2}, which have been computed as in Figure~\ref{fig:Tconst_stable}}
\label{fig:Tconst_unstable}
\end{figure}

Figures~\ref{fig:Tconst_stable},~\ref{fig:Tconst_unstable} refer to the case of constant transition probabilities discussed in Section~\ref{sect:alpha_beta.const}. In particular, in Figure~\ref{fig:Tconst_stable} where $\alpha>\alpha_\dagger$ we recover both the trends of the densities predicted by~\eqref{eq:Tconst.rho} and those of the mean viral loads predicted by~\eqref{eq:rho1m1}-\eqref{eq:rho2m2}. We notice, in particular, the decay to zero of the mean viral loads. Conversely, in Figure~\ref{fig:Tconst_unstable} where $\alpha<\alpha_\dagger$ we see that both the trends of the densities~\eqref{eq:Tconst.rho} and of the viral loads~\eqref{eq:rho1m1}-\eqref{eq:rho2m2} are still reproduced at the particle level but this time the mean viral loads blow as predicted by the qualitative analysis.

\begin{figure}[!t]
\centering
\subfigure[]{\includegraphics[width=.4\textwidth]{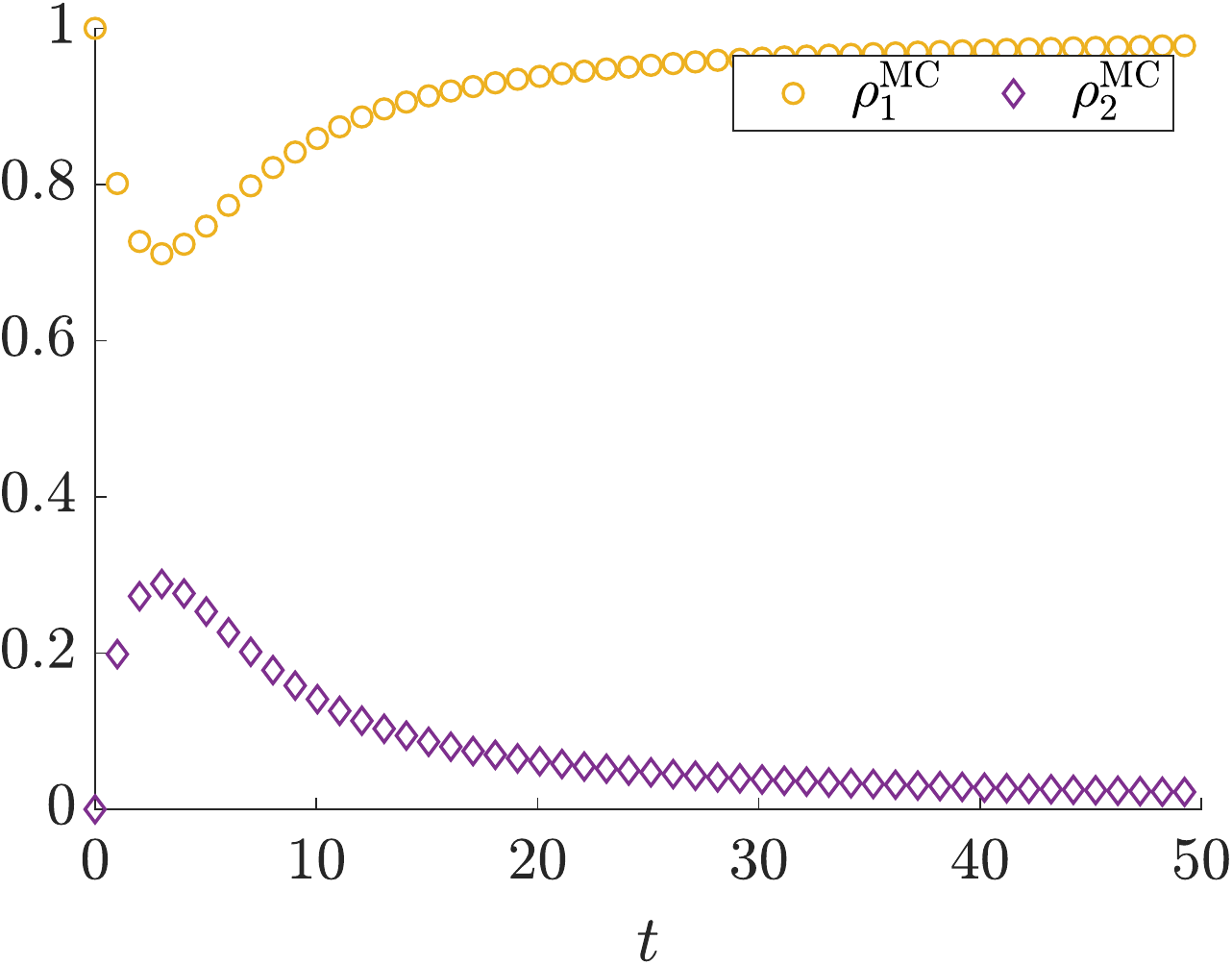}} \qquad
\subfigure[]{\includegraphics[width=.4\textwidth]{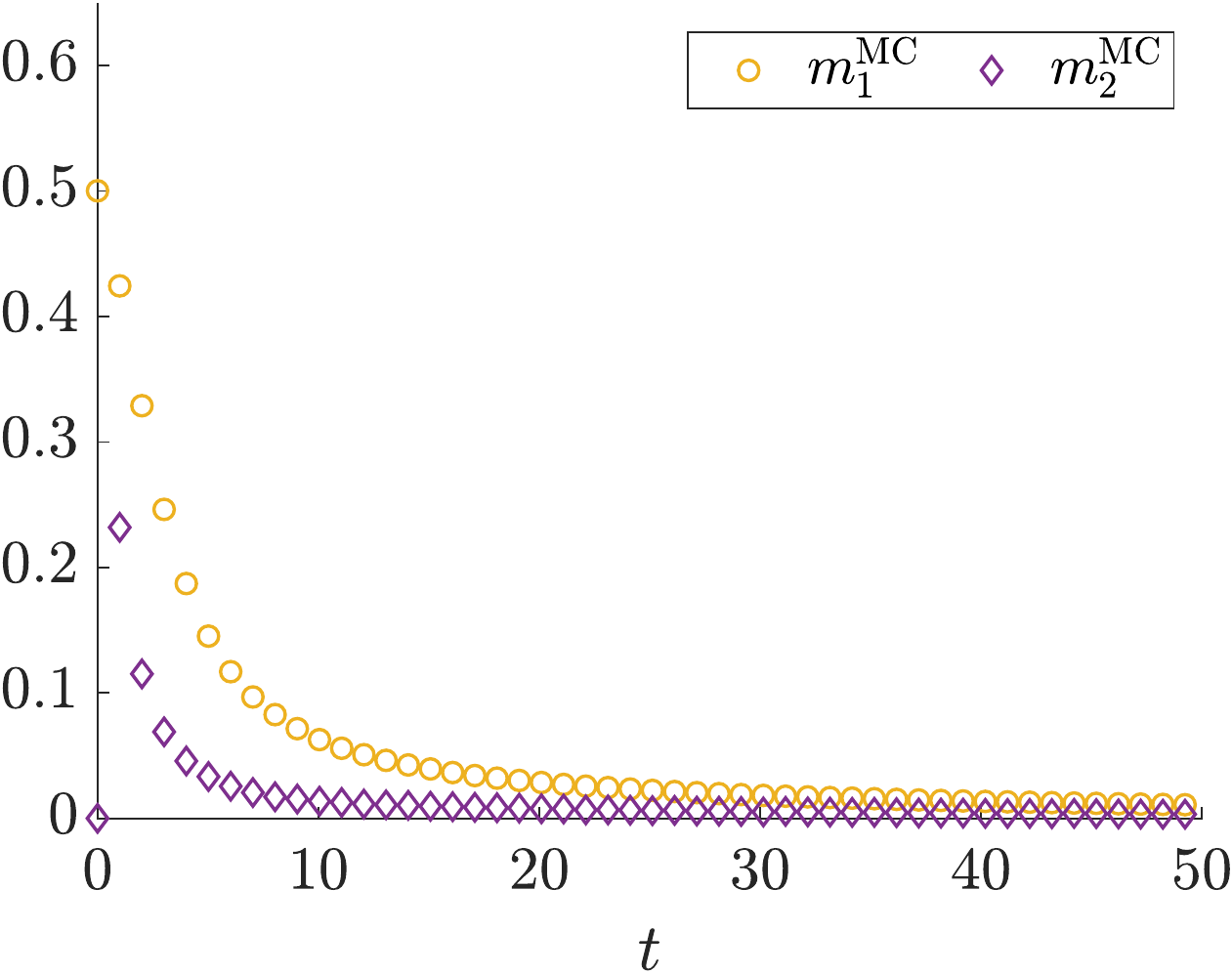}}
\caption{The problem of Section~\ref{sect:alpha_beta.var} with $\nu_1=\nu_2$ and $\mu\gg\lambda$}
\label{fig:Tvar_nu1=nu2_mu=10}
\end{figure}

\begin{figure}[!t]
\centering
\subfigure[]{\includegraphics[width=.4\textwidth]{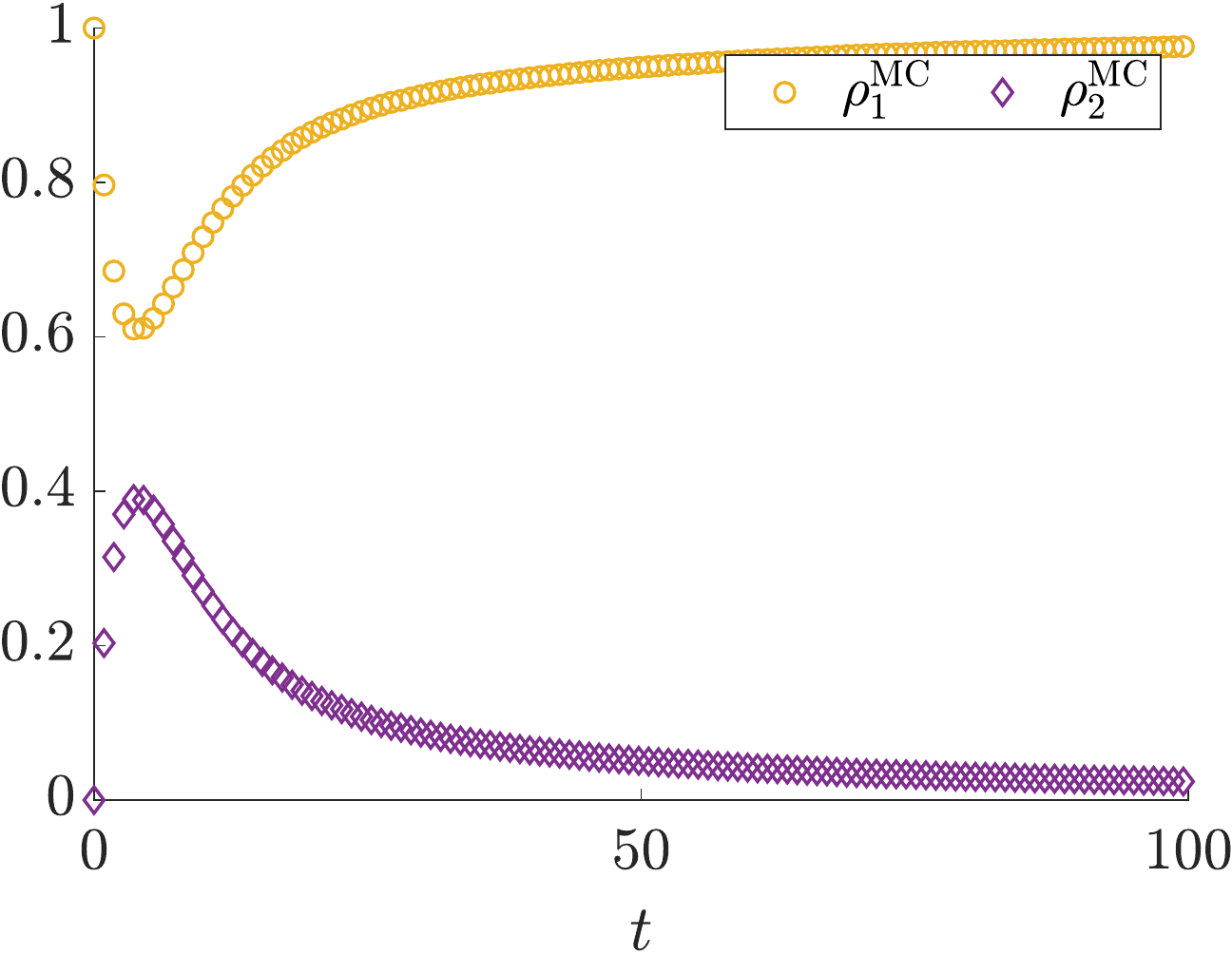}} \qquad
\subfigure[]{\includegraphics[width=.4\textwidth]{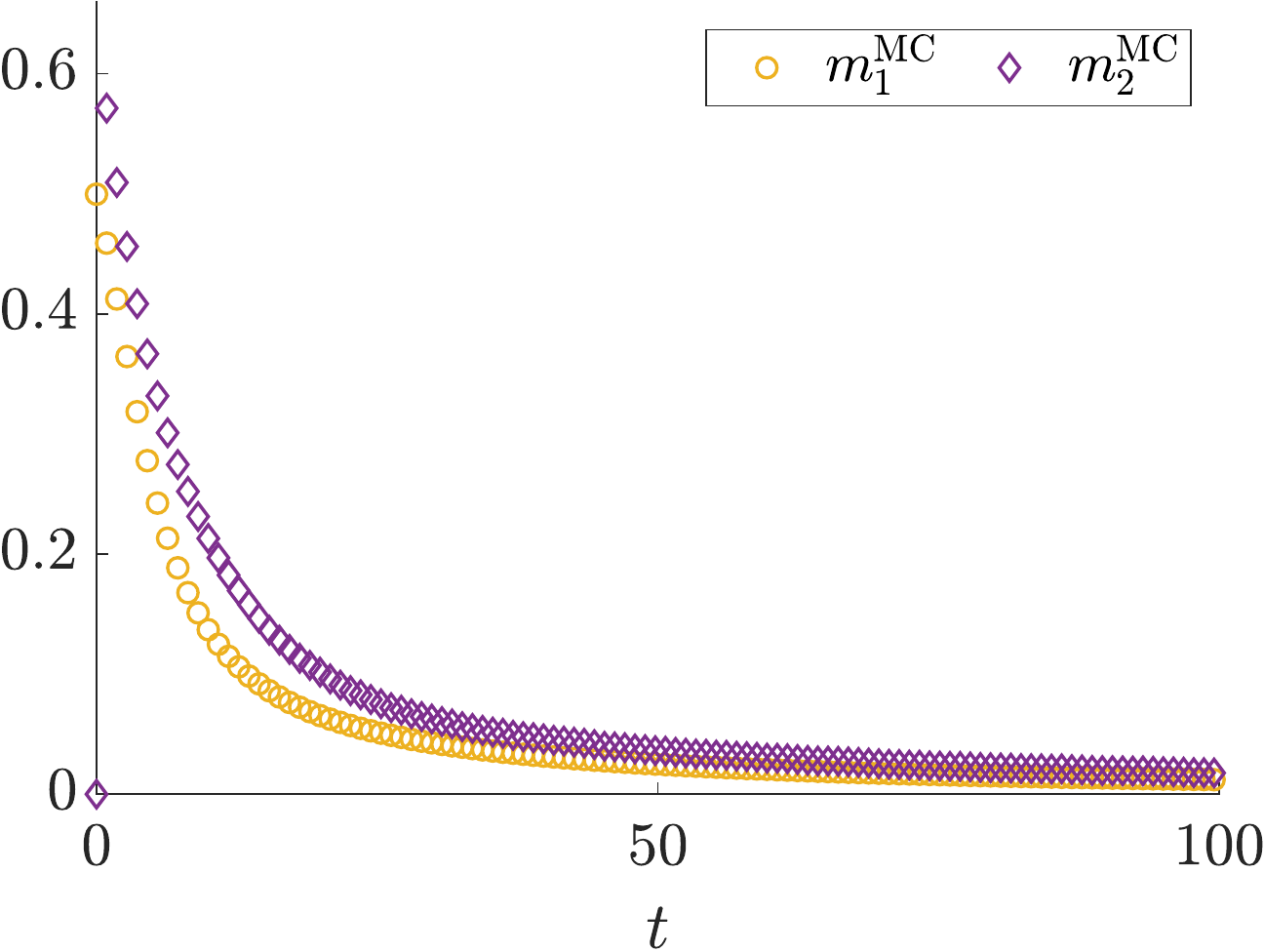}}
\caption{The problem of Section~\ref{sect:alpha_beta.var} with $\nu_1=\nu_2$ and $\mu=\lambda$}
\label{fig:Tvar_nu1=nu2_mu=1}
\end{figure}

\begin{figure}[!t]
\centering
\subfigure[]{\includegraphics[width=.4\textwidth]{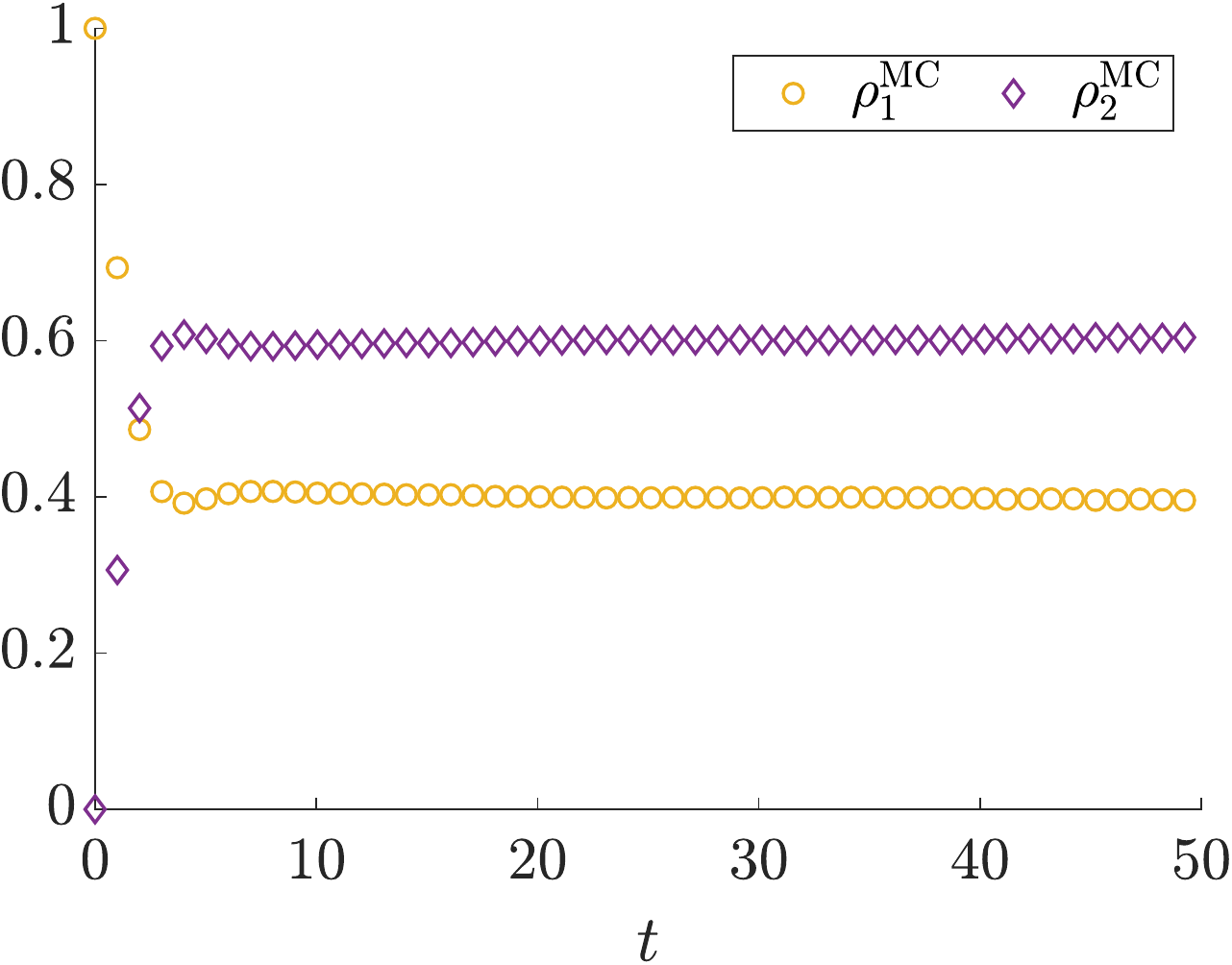}} \qquad
\subfigure[]{\includegraphics[width=.4\textwidth]{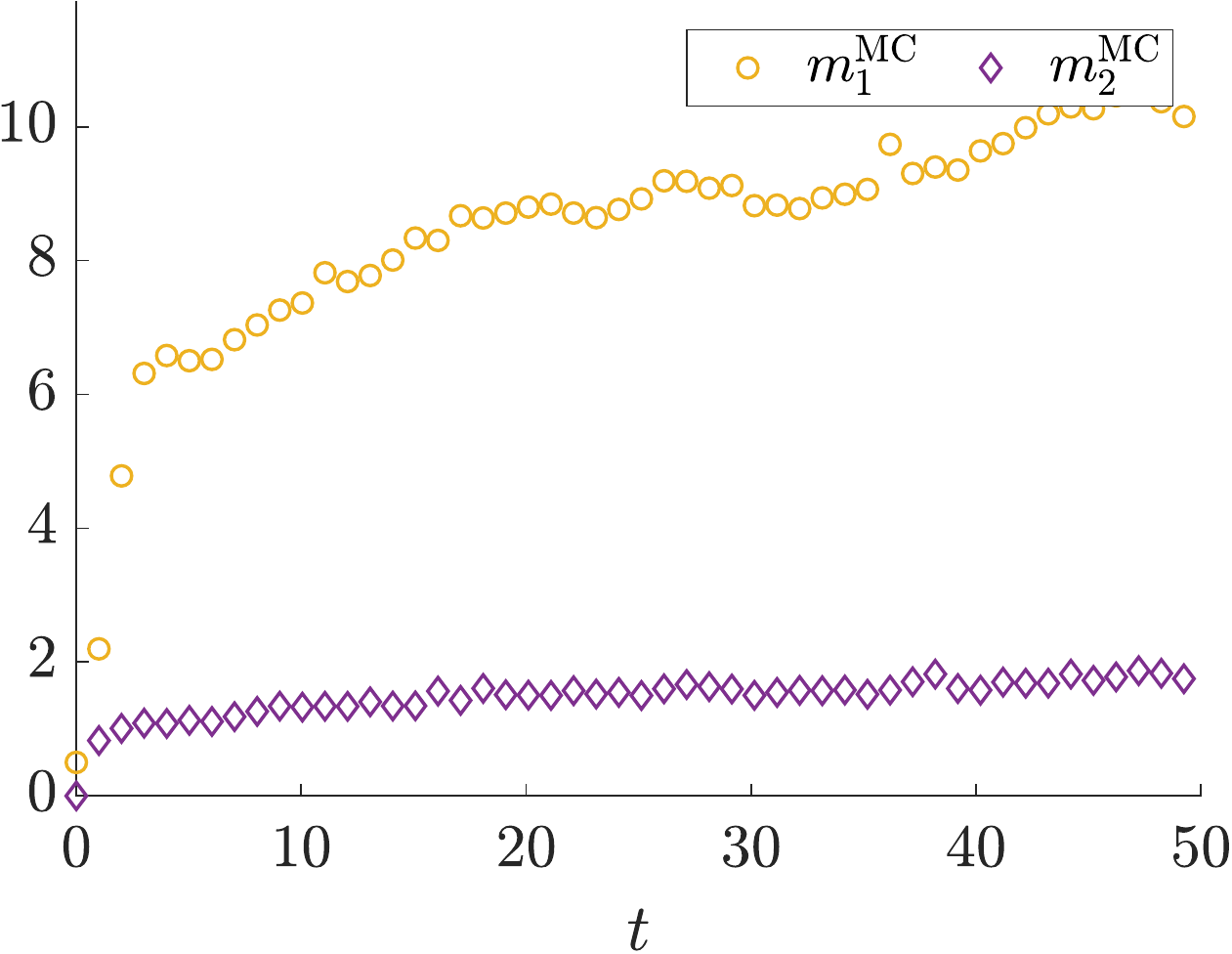}}
\caption{The problem of Section~\ref{sect:alpha_beta.var} with $\nu_1=0$ and $\mu\gg\lambda$}
\label{fig:Tvar_nu1=0_mu=10}
\end{figure}

\begin{figure}[!t]
\centering
\subfigure[]{\includegraphics[width=.4\textwidth]{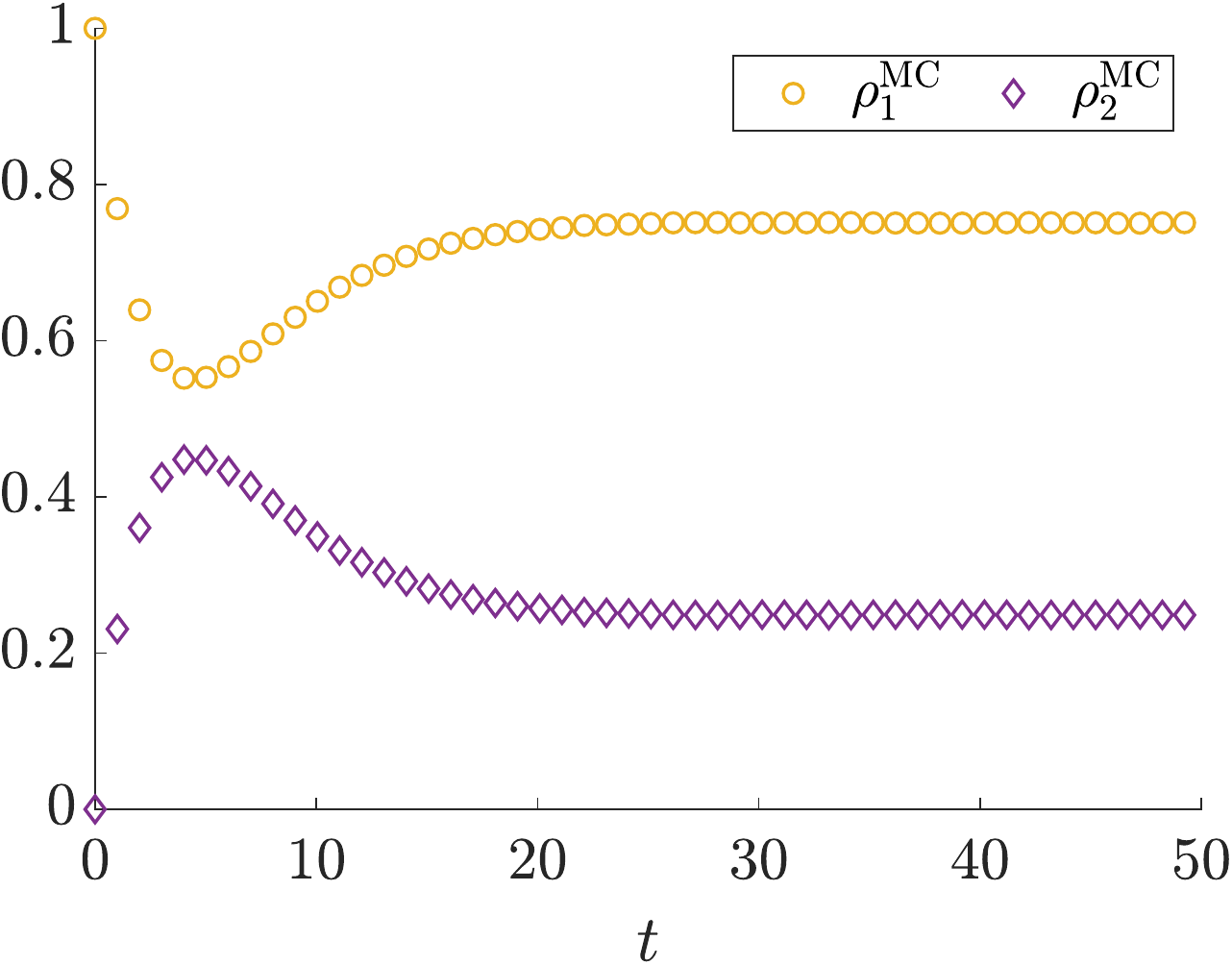}} \qquad
\subfigure[]{\includegraphics[width=.4\textwidth]{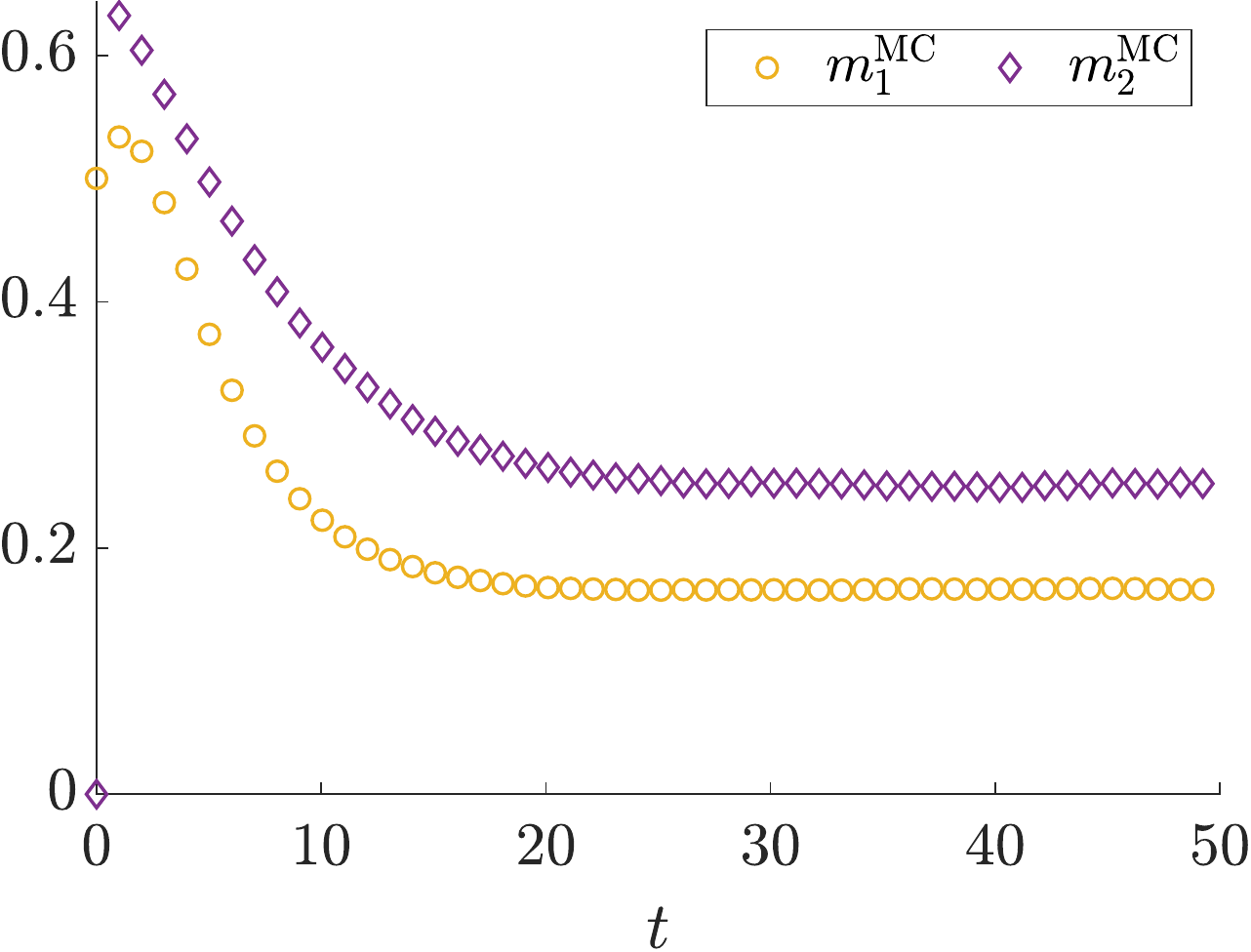}} \\
%\subfigure[]{\includegraphics[scale=0.7]{Tvar_nu1=0_mu=1-distr}}
\caption{The problem of Section~\ref{sect:alpha_beta.var} with $\nu_1=0$ and $\mu=\lambda$}
\label{fig:Tvar_nu1=0_mu=1}
\end{figure}

Figures from~\ref{fig:Tvar_nu1=nu2_mu=10} to~\ref{fig:Tvar_nu1=0_mu=1} refer instead to the case of variable transition probabilities discussed in Section~\ref{sect:alpha_beta.var}. Specifically, we set
$$ \alpha(v)\propto 1-e^{-v}, \qquad \beta(v)\propto e^{-v}, $$
which are respectively a monotonically increasing and a monotonically decreasing function with $\alpha(0)=0$ and $\beta(0)>0$. This way we reproduce exactly the conditions of the qualitative analysis of Section~\ref{sect:alpha_beta.var}.

In Figures~\ref{fig:Tvar_nu1=nu2_mu=10},~\ref{fig:Tvar_nu1=nu2_mu=1} we set $\nu_1=\nu_2$. Furthermore, in Figure~\ref{fig:Tvar_nu1=nu2_mu=10} we consider the regime $\mu\gg\lambda$, which produces a label switching-driven hydrodynamic evolution of the densities and mean viral loads based on a local-in-time equilibrium of the interactions. The Monte Carlo numerical solution confirms the theoretical predictions obtained in Section~\ref{sect:alpha_beta.var} by means of the hydrodynamic splitting~\eqref{eq:split.interaction.f1}-\eqref{eq:split.switching.f1} and~\eqref{eq:split.interaction.f2}-\eqref{eq:split.switching.f2}: in the long run, $\rho_1\to 1$ and $\rho_2\to 0$ with $m_1,\,m_2\to 0$. Conversely, in Figure~\ref{fig:Tvar_nu1=nu2_mu=1} we consider the regime $\mu=\lambda$, which does not allow for a hydrodynamic splitting of the kinetic equations because the interactions and the label switching take place on the same time scale. Although in Section~\ref{sect:alpha_beta.var} we have not explored this case, from the numerical results we observe that the qualitative trends of both the densities and the mean viral loads are very similar to those obtained for $\mu\gg\lambda$. In particular, up to a slightly slower rate of convergence in time, the asymptotic states are the same.

In Figures~\ref{fig:Tvar_nu1=0_mu=10},~\ref{fig:Tvar_nu1=0_mu=1} we finally examine the case $\nu_1=0$ in the frame of variable transition probabilities, that we have not investigated in Section~\ref{sect:alpha_beta.var}. This corresponds to infection dynamics~\eqref{eq:int.undiagnosed} such that an individual may only get more infected after coming into contact with another infected individual. In Figure~\ref{fig:Tvar_nu1=0_mu=10} we illustrate the case $\mu\gg\lambda$: since interactions are much quicker than label switches, the mean viral load of the undiagnosed individuals ($x=1$) tends to grow rapidly (Figure~\ref{fig:Tvar_nu1=0_mu=10}b). As a result, in the long run a large percentage of the population tends to be quarantined (Figure~\ref{fig:Tvar_nu1=0_mu=10}a). Finally, in Figure~\ref{fig:Tvar_nu1=0_mu=1} we illustrate the regime $\mu=\lambda$: this time, similarly to the case of Figure~\ref{fig:Tconst_stable}, the quarantine can control more effectively the spreading of the infection because the contagion among the individuals takes place on the same time scale of the label switches. Nevertheless, due to the infection-dependent transition probabilities, the infection is not completely eradicated in time but becomes endemic. In particular, Figure~\ref{fig:Tvar_nu1=0_mu=1}a shows that in the long run a fixed percentage (however lower than in Figure~\ref{fig:Tvar_nu1=0_mu=10}a) of individuals is systematically quarantined and Figure~\ref{fig:Tvar_nu1=0_mu=1}b confirms that such a quarantine is not fictitious like in the numerical test illustrated in Figure~\ref{fig:Tconst_stable}. Indeed, the quarantined population is not fully healthy like in Figure~\ref{fig:Tconst_stable}b because its asymptotic mean viral load is strictly positive. %Figure~\ref{fig:Tvar_nu1=0_mu=1}c show the normalised histograms which approximate numerically the asymptotic distributions $f_1^\infty,\,f_2^\infty$ and which illustrate these conclusions with an even greater statistical detail.

\section{Conclusions}
In this paper, we have considered Boltzmann-type kinetic models with label switching derived from stochastic microscopic dynamics accounting for the superposition of conservative interactions and group-wise non-conservative state-dependent relabelling of the agents. Remarkably, such a derivation has yielded straightforwardly a simple and efficient Monte Carlo particle scheme for the numerical approximation of the resulting kinetic equations.

For prototypical death and birth processes, we have been able to characterise explicitly both the transient and the equilibrium (``Maxwellian'') kinetic distributions in the special regime of sufficiently small parameters (quasi-invariant regime) by means of Fokker-Planck asymptotics and self-similar solutions.

Moreover, we have applied our kinetic framework to the construction of a simple, and certainly improvable, model of the contagion of infectious diseases with quarantine, which describes from a statistical mechanics point of view the interplay among:
\begin{enumerate*}[label=\roman*)]
\item the microscopic dynamics of contact and contagion among the individuals of a community;
\item the isolation of individuals diagnosed as infected;
\item the reintroduction in the community of quarantined individuals diagnosed as recovered.
\end{enumerate*}
In particular, the isolation and the reintroduction are regarded as label switches modelled on an viral load-dependent probabilistic basis. Thanks to its kinetic structure, this model depends on a relatively small number of parameters. Yet, it shows a quite rich variety of trends, which suggest clearly the impact of the microscopic features of the system on either the success or the failure of the quarantine as a control strategy of the global spreading of the infection. More importantly, the kinetic structure of the model has allowed us to address analytically several significant regimes by taking advantage of powerful methods of the kinetic theory, such as e.g., the hydrodynamic limit. This way, we have obtained a precise characterisation of the role of the microscopic parameters in the emergence of either global trend.

As research prospect, we mention that our kinetic equations with label switching provide a framework for the statistical modelling of \textit{network-structured} social interactions, see e.g.,~\cite{burger2020PREPRINT}, with the further possibility for the agents to jump from one node of the network to another. Applications include for instance social interactions on graphs, whose vertices represent spatial locations across which agents migrate or social compartments that the agents may change in time. Some of these applications are currently in preparation~\cite{dellamarca2021PREP,loy2021PREP} as developments of the model of the contagion of infectious diseases presented in Section~\ref{sect:contagion}.

\section*{Acknowledgements}
This research was partially supported  by the Italian Ministry for Education, University and Research (MIUR) through the ``Dipartimenti di Eccellenza'' Programme (2018-2022), Department of Mathematical Sciences ``G. L. Lagrange'', Politecnico di Torino (CUP: E11G18000350001) and through the PRIN 2017 project (No. 2017KKJP4X) ``Innovative numerical methods for evolutionary partial differential equations and applications''.

NL acknowledges support from ``Compagnia di San Paolo'' (Torino, Italy)

Both authors are members of GNFM (Gruppo Nazionale per la Fisica Matematica) of INdAM (Istituto Nazionale di Alta Matematica), Italy.

\bibliographystyle{plain}
\bibliography{LnTa-label_switching}

\appendix

\section{Numerical algorithm}
\label{app:nanbu}
\begin{algorithm}[H]
	\caption{Nanbu-Babovski algorithm with mass transfer for model~\eqref{eq:undiagnosed}-\eqref{eq:quarantined}}
	\label{alg:nanbu}
	\KwData{
		\begin{itemize}[noitemsep]
			\item $N\in\mathbb{N}$ total number of agents of the system;
			\item $N_1^n,\,N_2^n\in\mathbb{N}$ numbers of agents in $x=1$, $x=2$, respectively, at time $t^n:=n\Delta{t}$;
		\end{itemize}
	}
	Fix $\Delta{t}\leq\min\{\frac{1}{\lambda},\,\frac{1}{\mu}\}$\;
	\For{$n=0,\,1,\,2,\,\dots$}{
		Compute \newline{}
		$\rho_1^n=\dfrac{N_1^n}{N}, \qquad \rho_2^n=\dfrac{N_2^n}{N}, \qquad
			m_1^n=\dfrac{1}{N_1^n}\displaystyle{\sum_{k=1}^{N_1^n}}v_k^n, \qquad m_2^n=\dfrac{1}{N_2^n}\displaystyle{\sum_{k=1}^{N_2^n}}v_k^n$\;
		\Repeat{no unused pairs of agents are left}{
			Pick randomly two agents $(x_i^n,\,v_i^n)$, $(x_j^n,\,v_j^n)$ with $i\neq j$\;
			\For{$h=i,\,j$}{ \label{algo:label_switching-start}
				Sample $\Theta\sim\operatorname{Bernoulli}(\lambda\Delta{t})$\;
				\eIf{$\Theta=1$}{
					\If{$x_h^n=1$}{
						Sample $J\in\{1,\,2\}$ with law \newline{}
						$\P(J=1)=1-\alpha(v_h^n), \quad \P(J=2)=\alpha(v_h^n)$\;
					}
					\If{$x_h^n=2$}{
						Sample $J\in\{1,\,2\}$ with law \newline{}
						$\P(J=1)=\beta(v_h^n), \quad \P(J=2)=1-\beta(v_h^n)$\;
					}
					Set $x_h^{n+1}=J$\;
				}{
					Set $x_h^{n+1}=x_h^n$\;
				}
			} \label{algo:label_switching-end}
			Sample $\Xi\sim\operatorname{Bernoulli}(\mu\Delta{t})$\; \label{algo:interactions-start}
			\eIf{$\Xi=1$}{
				\If{$x_i^n=x_j^n=1$}{
					Update $v_i^n,\,v_j^n$ to $v_i^{n+1},\,v_j^{n+1}$ according to~\eqref{eq:int.undiagnosed}\;
				}
				\If{$x_i^n=1,\,x_j^n=2$ or vice versa}{
					Set $v_i^{n+1}=v_i^n$ and update $v_j^n$ to $v_j^{n+1}$ according to~\eqref{eq:int.quarantined} or vice versa\;
				}
				\If{$x_i^n=x_j^n=2$}{
					Update $v_i^n,\,v_j^n$ to $v_i^{n+1},\,v_j^{n+1}$ according to~\eqref{eq:int.quarantined}\;
				}
			}{
				Set $v_i^{n+1}=v_i^n$, $v_j^{n+1}=v_j^n$\;
			} \label{algo:interactions-end}
		}
	}
\end{algorithm}
\end{document}